\documentclass[preprint,aps,amsmath,superscriptaddress,nofootinbib,tightenlines
]{revtex4}
\usepackage{axodraw,pstricks,color}
\usepackage{bm}
\usepackage{epsfig}

\def\Slash#1{{#1\!\!\!\slash}}

\def\nslash{n\!\!\!\slash}
\def\bnslash{\bar n\!\!\!\slash}
\def\pslash{p\!\!\!\slash}

\def\OMIT#1{}

\newcommand{\nn}{\nonumber} 

\newcommand{\bn}{{\bar n}}
\newcommand{\bea}{\begin{eqnarray}}
\newcommand{\eea}{\end{eqnarray}}
\newcommand{\bnp}{\bar n \!\cdot\! p}

\newcommand{\cP}{{\cal P}}

\newcommand{\cC}{ {\mathrm C} }
\newcommand{\cO}{{\mathcal O}}
\newcommand{\cJ}{{\mathcal J}}

\newcommand{\cA}{{\mathcal A}}

\newcommand{\cX}{{\Omega}}
\renewcommand{\cP}{{\mathcal P}}

\newcommand{\psl}{{\pslash}}
\newcommand{\qsl}{{\qslash}}
\newcommand{\nsl}{{\nslash}}
\newcommand{\nbsl}{{\bnslash}}
\newcommand{\qbar}{{\bar{q}}}
\newcommand{\nbar}{{\bar{n}}}
\newcommand{\bfn}{{ \mathbf{n} }}
\def\Asl{A\!\!\!\slash}
\def\cAsl{{\cal A}\!\!\!\slash}

\def\gsl{g\!\!\!\slash}
\def\pperpsl{p\!\!\!\slash_\perp}
\def\rd{{\mathrm{d}}}
\newcommand{\pq}{p_q}
\newcommand{\pqb}{p_\qbar}
\newcommand{\pg}{p_g}
\def\qsl{p\!\!\!\slash_q}
\def\qbsl{p\!\!\!\slash_\qbar}
\def\gsl{p\!\!\!\slash_g}
\newcommand{\mut}{\mu}
\newcommand{\msbar}{ \overline{\mathrm{MS}} }
\newcommand{\qcd}{{ \cJ }}
\newcommand{\scet}{{ {\cO } }}
\newcommand{\tdot}{\!\cdot\!}
\newcommand{\QQ}{ Q^2}
\newcommand{\qqqq}{ q \qbar q \qbar}
\def\ccol{}
\def\dcol{}
\newcommand{\cusp}{ {\Gamma} }
\newcommand{\dusp}{ {B} }
\newcommand{\rem}{ \Omega }
\newcommand{\emi}{ {\mathcal{E}}^\mu} 
\newcommand{\emidag}{ {\mathcal{E}}^\dagger_\mu } 
\newcommand{\eps}{ {\varepsilon} }
\newcommand{\ycut}{y_{\mathrm{cut}} }
\newcommand{\ircut}{ \Lambda_{\mathrm{IR}}}
\newcommand{\sep}{\sigma_3^\eps }
\newcommand{\sbep}{\sigma_2^\eps }

\begin{document}


\preprint{\vbox{ \hbox{xxx-xxx-xxx} }}

\title{Event generation from effective field theory} 

\author{Christian W.~Bauer\footnote{Electronic address: cwbauer@lbl.gov}}
  \affiliation{Ernest Orlando Lawrence Berkeley National Laboratory and
University of California, Berkeley, CA 94720}

\author{Matthew D.~Schwartz \footnote{Electronic address: mdschwartz@lbl.gov}}
  \affiliation{Ernest Orlando Lawrence Berkeley National Laboratory and
University of California, Berkeley, CA 94720}
\date{\today\\ \vspace{1cm} }


\begin{abstract}
A procedure is developed for using Soft Collinear Effective Theory (SCET) to generate
fully exclusive events, which can then be compared to data from collider experiments.
We show that SCET smoothly interpolates between QCD for hard emissions, and the parton
shower for soft emissions, while resumming all large logarithms.
In SCET, logarithms are resummed using the renormalization
group, instead of classical Sudakov factors, so subleading logarithms can be resummed as well.
In addition, all loop effects of QCD can be reproduced in SCET, which allows the effective theory
to incorporate next-to-leading and higher-order effects.
We also show through SCET that in the soft/collinear limit,
successive branchings factorize, a fact which is essential to parton showers, and that
the splitting functions of QCD are reproduced. Finally, combining these results, we present a example
of an algorithm that incorporates the SCET results into an event generator which is systematically
improvable.
\end{abstract}

\maketitle

\newpage

\section{Introduction}
To test a model of particle physics we must be able to describe the distribution
of particles it predicts. Only a fully exclusive distribution is truly useful, 
since in order to obtain 
a meaningful comparison between theory and experiment, the theory
must undergo a detector simulation and be subjected 
to the same cuts as the experimental data.
However, 
high energy colliders produce events with thousands of particles in each event. 
There is no easy way to describe even the phase space 
for such complicated final states, and therefore Monte Carlo simulations
have become essential for the analysis of every high energy experiment. 
Practically speaking,
the current generation of Monte Carlo tools has been able to reproduce the standard model remarkably well. But with
the onset of the LHC, new energy scales and new kinematical configurations, such as events with many high $p_T$ jets, may
appear, and these tools will be pushed beyond their validity.
So it is our task to help these simulations
incorporate as much information from analytical calculations as possible, including loop corrections and the cancellation of
infrared divergences when appropriate, while still having them produce exclusive events. 

The problem with current techniques is that a number of 
not necessarily good approximations are forced by practical considerations.
For example, it is unreasonable to calculate the analytic expression for a thousand-particle amplitude, and so
event generators resort to the parton shower approximation. This approximation starts with an event
resulting in, say, two quarks. These quarks then branch into quarks and gluons, and evolve down in
energy until they hadronize at some low infrared (IR) scale. The branching is treated as a classical Markov process
with emissions governed by splitting functions derived in the strict collinear limit;
thus any quantum mechanical interference effects are lost.  
A lot of work has gone into improving the results from Monte Carlo simulations. Mostly, it has been
directed towards incorporating higher order QCD effects to improve the distributions in regions where
the parton shower cannot be trusted. And, generally, quite good agreement with data has been achieved.
However, it remains an extremely important open question to estimate the errors in these techniques,
and to be able to reduce those errors systematically. It goes without saying that comparing theoretical predictions to
data, or comparing the output of different Monte Carlo schemes, 
is not the ideal method of estimating error when searching for new physics.

Part of the difficulty in simulating QCD is that it becomes strongly coupled at large distances. 
But even when QCD is weakly
coupled, there is not an obvious perturbation expansion. Of course, we can expand a differential cross
section as
\begin{equation}
\rd \sigma = \sum_n \left( \frac{\alpha_s}{\pi}\right)^n \rd \sigma^{(n)}\,.
\end{equation}
But when there are multiple scales in the event, 
such as the relative momentum of pairs of final state partons, large logarithms may appear at any order.
Typically,
\begin{equation}
\rd \sigma^{(n)} = \sum_{m < 2n} c_m^{(n)} \log^m \frac{p_i}{Q}\,,
\end{equation}
where $p_i$ is some kinematical variable, and $Q$ is a fixed reference scale, such as the center-of-mass
energy of the collision. Even if $\alpha_s \ll 1$ we may have $\alpha_s \log^2 (p/Q)\sim 1$ and
so the perturbation expansion breaks down. In this case, the large logarithms need to  be resummed.  Parton showers do
this resummation, but without the right framework, it is easy to lose track of which
terms are accounted for and which are not. The right framework is an effective field theory which 
can sum the logarithms, incorporate finite $\alpha_s$ corrections, and be extended with power
corrections to fully reproduce QCD to any desired accuracy. Such an effective theory
description was introduced in~\cite{BS}, and in this work we will give more details
and expand on that approach.

An effective theory for event generation should not disturb the infrared properties of QCD,
where strong dynamics and hadronization take place. 
Infrared divergences in QCD, such as a logarithmic
dependence on the jet resolution scale, can be physical and must be reproduced. These divergences are
either soft, for example, when a gluon's energy goes to zero, or collinear, such as when a quark emits
a possibly hard gluon in almost the same direction. 
The effective theory with all these properties is the Soft Collinear Effective Theory
(SCET)~\cite{SCET1,SCET2,SCET3,SCET4}.
In this paper, we show how SCET can be implemented in an event generator so that the results are
systematically improvable. 

We will demonstrate how at leading order, SCET is equivalent to the 
traditional parton shower approach. Parton showers write the probability for a branching 
as
\begin{equation}
\rd \sigma = \Delta(\tau) P(\tau,z) 
\,,
\end{equation}
where $P(\tau,z)$ is a splitting function and $\Delta(\tau)$ a Sudakov factor
\begin{equation}
\Delta (\tau)
= \exp\left( - \int  \! dz \, P(\tau,z) \right)\,.
\end{equation}
The splitting function represents the probability of a parton to split, 
and the Sudakov factor accounts for the fact that if a parton splits at a scale $\tau$,
it should not have already split. We will see the splitting function derived
from collinear emissions in SCET and the Sudakov factor reproduced from renormalization group (RG) evolution.
One of the important facts of QCD, which gives rise to parton showers and which allows Monte Carlo techniques simulate multiple branchings, is that a given cross section factors into a 
product of probabilities, namely the probability to create an initial final state, 
multiplied by the probabilities to have subsequent splittings. This will rederived from SCET.

Besides justifying the parton shower approach, SCET can be used to improve it.
The problem of how to properly combine QCD matrix elements with parton showers is
first cast in the language of scale separation. 
When the logarithms appearing in the differential cross section are large, the
scales are widely separated, and the effective theory can be matched at the higher energy scale,
and run down to the lower scale. The splitting functions used in the parton shower describe the long distance 
behavior of QCD (and of SCET), while the short distance physics is different in the two theories.
Although the short distance physics of QCD is not fundamentally
part of SCET, it can be fully reproduced through a consistent matching procedure. Thus SCET is valid at all
scales, in contrast to QCD or the parton shower separately.

We would like this paper to be self-contained and generally 
readable by both SCET and the Monte Carlo communities,
as well as people familiar with neither field. So we attempt to incorporate
a terse and incomplete review of event generators in Section~\ref{sec:eventgenerators}
and of SCET at the beginning of Section~\ref{sec:details}
and in Appendix~\ref{app:scet}. 
Section~\ref{sec:schematic}  gives a schematic presentation 
of the general idea of event generation in SCET. 
As much as possible, we sequester
the detailed calculations to Section~\ref{sec:details}, which is the heart of the paper. 
The results of Section~\ref{sec:details} are applied in various ways in 
Section~\ref{sec:comp}. We show first how Sudakov factors are reproduced from the renormalization
group in SCET. A discussion is included of next-to-leading log resummation. We then discuss how
the splitting functions and the factorization of successive branchings are understood from SCET.
Next, we explore NLO effects, which include the cancellation of infrared divergences in physical observables.
Finally, we use the SCET results to calculate some observables from parton-level results.
We present the thrust distribution for 3-partons states, and the 2-jet fraction $\sigma_2$.
Section~\ref{sec:algorithm} first summarizes how differential cross-sections are calculated, and then
describes a simple algorithm that can use these cross section to distribute events in a Monte Carlo program.
Finally we present our conclusions and outlook. We also include in Appendix~\ref{app:kin}
some kinematical relations and conventions that are used throughout the paper.

\section{Introduction to parton showers and event generators}
\label{sec:eventgenerators}

In this section we review some aspects of how parton showers and event generators work,
with an eye towards comparing to the SCET approach. 
This section contains no new information, and readers 
familiar with the subject can safely skip this section.

An event is typically
generated in three phases~\cite{pythia,herwig,sherpa,Dobbs:2004qw}. 
First a simple hard process is selected at the parton level, 
with a probability proportional to its production cross section calculated using standard 
Feynman diagram methods. 
Second, the partons, which
are taken to be highly off-shell at the hard scale, radiate additional partons and ``evolve'' 
down
until their off-shellness reaches a low scale, which is typically chosen to be of order 1 GeV. Finally, the partons hadronize using some confinement model.
We will be concerned here mainly with the second stage, the parton shower, and its interplay with the other two.

Parton showers use classical evolution to describe the emissions of particles:
they assign a probability for one particle to split in two.
This probability depends on two kinematical variables. 
The first variable, $\tau$, is chosen to be a measure of the virtuality of the initial particle, and tends to zero if the two final
particles are collinear. The second variable, $z$, measures the
relative energy between the two final particles.  
Using that the splitting is independent of azimuthal angle, 
and that the branching probability is linearly divergent as $\tau$ goes to zero,
the probability is proportional to 
\begin{eqnarray}
P(\tau,z) = \frac{\alpha_s}{2 \pi} \frac{P_{q g}(z)}{\tau} \label{Pdef}
\end{eqnarray}
up to terms of order $\tau^0$. 
Here, $P(\tau,z)$ is our notation, while $P_{a b}$ is the traditional spin averaged splitting function, 
with $a,b$ indexing the final state particles.
For example, the quark splitting function in QCD is
\begin{equation}
P_{q g}(z) = C_F \frac{1+z^2}{1-z} \label{splitfn}
\end{equation}
where $C_F=\frac{4}{3}$.

In order to calculate the classical probability of an initial particle with 
virtuality $\tau_1$ to branch at a particular value $\tau_2$, one needs to
include the probability that no branching has occurred at a larger value of $\tau$. 
This is similar to the well known case of nuclear beta decay, where this no-branching
probability is responsible for the exponential decay rate. 
The probability
of no-emission from $\tau_1$ to $\tau_2$ is then given by an integral, known as a Sudakov factor 
\begin{equation}
\Delta(\tau_2,\tau_1) = \exp\left(-\int_{\tau_1}^{\tau_2} 
\frac{\rd\tau}{\tau} \int \rd z \frac{\alpha_s(\mu[\tau,z])}{2 \pi} P_{ab}(z)\right)
 \label{sud}
\end{equation}
Here, the scale $\mu$ at which $\alpha_s$ is evaluated can depend on both $\tau$ and $z$. 

As an example, consider a quark-antiquark pair with virtuality $Q$. 
The probability that a branching occurs at a scale $\tau$ is given by the probability 
that neither quark branched at a scale greater than 
$\tau$, times the probability for either of the two to branch at that scale. Thus the differential 
cross section to have three final state partons is given by the cross section for two final state 
particles, multiplied by a Sudakov factor and the sum of the two splitting functions for quark and 
antiquark emission
\begin{equation}
{\mathrm{d}} \sigma_3 = {\mathrm{d}} \sigma_2\, \Delta(Q,\tau)^2 \left[P(\tau,z) + P'(\tau,z)\right]\,.
\label{dsps}
\end{equation}
where $P'$ is the equivalent of $P$ for the antiquark emission.

Parton showers rely on two crucial assumptions. First, 
one neglects the interference between the emissions off the various particles, and 
second, the intermediate states are taken onshell when deriving the splitting 
functions. For example, in the emission of a gluon off a 
quark-antiquark pair, the square of the full matrix element in QCD is replaced by
a sum of two independent emission graphs
\begin{equation}
\SetScale{0.5}
\left|
\fcolorbox{white}{white}{
\begin{picture}(60,10) (0,3)
  \SetWidth{0.5}
    \SetColor{Black}
    \Line(60,8)(120,8)
    \Line(0,8)(60,8)
    \COval(60,8)(6.32,6.32)(-161.56505){Black}{White}
    \Line(61.41,12.24)(58.59,3.76)
    \Line(64.24,6.59)(55.76,9.41)
    \Gluon(45,8)(0,23){5}{3.44}
\end{picture}
}
+
\fcolorbox{white}{white}{
\begin{picture}(60,10) (0,3)
    \SetWidth{0.5}
    \SetColor{Black}
    \Gluon(75,10)(120,25){5}{3.44}
    \Line(60,10)(120,10)
    \Line(0,10)(60,10)
    \COval(60,10)(6.32,6.32)(-161.56505){Black}{White}
    \Line(61.41,14.24)(58.59,5.76)
    \Line(64.24,8.59)(55.76,11.41)
\end{picture}
}
 \right|^2 
\to 
\left|
\fcolorbox{white}{white}{
  \begin{picture}(40,10) (0,-2)
    \SetWidth{0.5}
    \SetColor{Black}
    \Line(0,-1)(45,-1)
    \Line(45,-1)(90,-1)
    \COval(45,-1)(6.32,6.32)(-161.56505){Black}{White}
    \Line(46.41,3.24)(43.59,-5.24)
    \Line(49.24,-2.41)(40.76,0.41)
  \end{picture}
}
\right|^2
\left(
\left|
\fcolorbox{white}{white}{
  \begin{picture}(30,10) (0,1)
    \SetWidth{0.5}
    \SetColor{Black}
    \Line(0,5)(60,5)
    \Gluon(45,5)(0,20){5}{3.44}
  \end{picture}
}
\right|^2
+
\left|
\fcolorbox{white}{white}{
\begin{picture}(30,10) (0,1)
    \SetWidth{0.5}
    \SetColor{Black}
    \Line(0,5)(60,5)
    \Gluon(60,20)(15,5){5}{3.44}
  \end{picture}
}
\right|^2
\right)\,.
\end{equation}
Both of these assumptions can be justified in the limit $\tau \to 0$, which implies that the virtuality
of the branching particle is small compared to its energy. In that limit the branching particle becomes almost on-shell and the interference between the two QCD diagrams becomes subdominant.

Note that the two diagrams on the right hand side are not gauge invariant and 
thus not well-defined. 
The parton shower circumvents this problem by summing only over physical polarizations, 
which restores gauge invariance, {\it{ipso facto}}. The splitting functions are 
well defined because the residue of the pole is gauge invariant and does not get a contribution
from interference. But the $\tau^0$ and higher order pieces depend on conventions. For
example, take the square of the relative transverse momentum between the two final 
state particles, $p_T^2$,  as the 
measure of the virtuality $\tau$. The branching probability is then proportional to $P(\tau,z)$
 with $\tau = p_T^2$. Now consider the different choice
$\tau = t$, where $t$ denotes the invariant mass of the two final state particles $t = (p_i + p_j)^2/Q^2$. 
These two variables are related by
\begin{equation}
\label{pT-t_rel}
p_T^2 = \frac{t(z-t+tz)(1-z-tz)}{(t-1)^2} = tz(z-1) + {\cal O}(t^2)
\end{equation}
Using this to change variables in $P(t,z)$, we reproduce the same pole terms 
as in $P(p_T^2,z)$, however the higher order terms differ. Thus we cannot define the higher
order terms in splitting functions in a consistent way.

From the discussion so far we have learned that parton showers give 
simple expressions for the emission of partons. These 
can easily be turned into powerful computer algorithms using Monte Carlo techniques, which 
can be used to generate final states with many partons. 
One starts with a cross section for a process with a limited number of 
particles in the final state. In practice, event generators often 
start with processes with only two final state partons. The virtuality of these two partons 
is chosen to be comparable to the hard scale of the interaction $Q$, and all 
additional partons in the final state are generated by the classical probabilities of 
particles to split into two particles with lower virtuality. This can 
be cast in a Markov Chain process, evolving the 
system from high to low virtuality, adding additional particles through 
the splitting functions. 

Different parton shower algorithms all use the physics described above, but they differ in 
which next-to-leading order effects they incorporate. 
First, they use different choices for the evolution variable $\tau$~\cite{Mrenna:2003if,Gieseke:2003rz}. Second, 
they use different choices for the scale $\mu$ at which
$\alpha_s(\mu)$ is evaluated in \eqref{sud}. Finally, for each emission in a parton shower, 
$z$ and $\tau$ are distributed over phase space assuming that the partons are on shell. 
But subsequent branchings require
the partons to be off shell, so there is an ambiguity in how to assign the
 kinematics~\cite{Catani:1996vz}.
 All of these 
effects only contribute at next-to-leading
order, but they can give rise to considerable differences between 
the different programs in certain cases.

Another aspect of QCD that has to be taken into account carefully arises for 
the emission of a soft parton~\cite{Bengtsson:1986et,Bengtsson:1986hr}. 
The analog of the Chudakov effect from
QED is that, when integrated over azimuth, 
soft emissions at large angle are sensitive only to the aggregate color 
charge of all the contributing partons. The result is
that large angle soft emissions are suppressed. 
This is generally incorporated into parton showers by either using an
evolution variable which corresponds to angle (as in Herwig~\cite{herwig})
or by explicitly vetoing emissions which are not angular ordered (as in some versions of
Pythia~\cite{pythianew}).

The assumptions of parton showers restrict their 
validity to regions of phase space where each successive emission has to have 
virtuality much smaller than any previous one, $Q \gg \tau_1 \gg \tau_2 \gg \ldots\,.$
Regions of phase space with $\tau \sim Q$ are only correctly described by the 
full QCD matrix elements. Thus, only an event generator which combines
both the full QCD matrix elements with the parton showers will yield predictions
which are correct to a given fixed order in perturbation theory, while also 
summing the leading logarithms correctly and allowing for the additional production
of partons with small transverse momenta.
Several techniques have been developed which allow such a combination of QCD matrix
elements with parton showers, including original work which has
been implemented in Pythia and 
Herwig~\cite{Seymour:1994df,Miu:1998ju,Norrbin:2000uu}
and some more recent developments~\cite{CKKW,Mrenna:2003if,Krauss:2002up,Schalicke:2005nv,Hoche:2006ph,Lonnblad:2001iq}. 
One popular example is the CKKW procedure~\cite{CKKW},
which provides a way to combine tree level QCD
matrix elements with parton shower evolution.
The idea is to compute the exact tree-level QCD matrix element for
an event, including interference, but with a lower cutoff $\tau_0$ on the virtuality between 
any two particles. One then reconstructs the dominant diagram using the $k_T$ 
algorithm~\cite{kt}
and reweights the event by multiplying with appropriate Sudakov factors. 
Parton showers are then added to these matrix elements, vetoing all emissions
with $\tau > \tau_0$. This avoids double counting between the emissions contained
in the QCD matrix elements and those generated by the parton showers. 
This algorithm has been tested against data, 
and overall the procedure seems to work very well~\cite{Schalicke:2005nv}.

A truly accurate next-to-leading order calculation would also incorporate
loop diagrams.
To obtain the matrix elements at NLO requires 
calculating the one loop 
correction to the lowest order matrix element and combining it with the 
matrix element which
describes the radiation of one additional parton. The difficulty in 
going to NLO is that both the
virtual contributions at one loop and the real emission have infrared 
divergences, which only 
cancel when both diagrams are combined to calculate an infrared safe 
observable~\cite{Kinoshita:1962ur,Lee:1964is}, and
this cancellation is difficulty to encode numerically.
MC@NLO~\cite{MCNLO} provides one solution (see 
also~\cite{Dobbs:2001dq,Potter:2001ej,
Collins:2000gd,Kramer:2003jk,Soper:2003ya,Nagy:2005aa,Davatz:2006ut}). It
uses the fact that the first splitting in the parton shower reproduces 
the IR divergence of the
real QCD emission, and can thus be used to devise a subtraction from 
both the real and virtual
diagrams that render both these contributions finite. This yields two 
separate matrix elements, which
are both finite and can be used as starting conditions for a traditional parton shower. 
The initial results appear promising~\cite{Frixione:2003ei}. One problem is that
MC@NLO works at the level of cross sections, not matrix elements, so it runs up 
against the possibility of having negative weights. It is also not clear how to generalize
the procedure to higher orders.

\section{Jet distributions from SCET: Schematics}
\label{sec:schematic}
As we have seen in the previous section, the traditional Monte Carlo method uses
splitting functions and Sudakov factors to generate a fully showered event.
An event generator typically starts with an underlying hard process calculated using matrix elements 
of the full theory. 
The splitting functions then generate additional partons from these 
simple final states.
The Sudakov factor, which is the probability of no-branching, is included,
and it resums the large logarithms at leading order. 

The traditional Monte Carlo method uses
splitting functions and Sudakov factors to generate a fully showered event. Both splitting functions and Sudakov factors can be derived in the limit of small transverse momentum, and parton showers are only correct in the limit
where each successive branching has $p_T$ much smaller that any previous 
branching: $Q \gg p_T^{(1)} \gg p_T^{(2)} \ldots \,.$ Thus, only events with widely separated 
momentum scales can be described using parton shower techniques. 

The occurrence of these 
widely separated scales makes it natural to reformulate the problem 
in the language of effective field theory. The appropriate effective theory for this problem is the soft-collinear effective
theory (SCET), which is designed to reproduce exactly 
long distance physics in the limit of collinear or soft 
radiation. 
SCET is a simplified version of QCD where only collinear and soft degrees of freedom are kept and all others 
have been integrated out. Any calculation using effective field theories 
requires three steps. First, one needs to match the
effective theory onto the underlying theory. Matching determines 
coefficients in the effective theory such  
that at some short distance scale it exactly reproduces the underlying 
theory. This ensures that the effective theory below that scale contains
the same information as the underlying theory. The scale for 
this matching calculation is typically chosen to coincide with the hard
scale in the problem, for example the center of mass energy of
the collision. This ensures that the resulting 
short distance coefficients do not contain any large logarithms. 
Second, the effective  theory is evolved to the lower scales 
arising in  the process one wants to describe. This is achieved 
technically using 
renormalization group (RG) evolution, which sums logarithmic terms of the ratio 
of the low to the high scale. Finally, the matrix elements of the operators 
in the effective theory are calculated at the low scale.

As we will show in this paper, these steps naturally correspond 
to the ingredients in traditional parton showers mentioned above. 
The matching calculation encodes the hard underlying 
process which we want to study. The solution to the RG evolution gives 
rise to evolution kernels, which are equivalent to appropriate 
combinations of Sudakov factors. And the resulting matrix elements in 
SCET have the property that in the collinear limit their squares 
simplify to squares of  simpler operators, multiplied by the 
splitting functions of QCD.

The first step is to match the full theory onto the effective theory 
at the hard scale $Q$. To match, 
we introduce operators in SCET and choose their coefficients such that the matrix elements in
QCD and SCET are the same. 
The matching can be done perturbatively in $\alpha_s$ to any order.
Once we have matched at the hard scale, we no longer need QCD. To be more specific, consider the process $e^+e^- \to $ hadrons. In the standard model, this is mediated by a current
\begin{equation}
\cJ = \qbar \Gamma q \label{qcdj}\,,
\end{equation}
where $\Gamma = \gamma^\mu$ in the case of an intermediate photon or
$\Gamma = g_V \gamma^\mu + g_A \gamma^\mu \gamma^5$ if the $Z$ boson is included.
The current has a non-vanishing matrix element in many final states, and we can compute
$\langle \cJ| p_1\cdots p_n\rangle \equiv \langle 0 | \cJ | p_1\cdots p_n\rangle$ for any number of partons.
In SCET, the process is mediated by operators which are  constructed out of the fundamental
objects of SCET: soft and collinear gluons, collinear quarks, and Wilson lines.
Wilson lines are required to ensure gauge invariance of SCET, and each 
collinear field needs to be multiplied with Wilson lines in the appropriate 
SU(3) representation. For example, collinear fermions and collinear gluons
always come in the combinations
\begin{eqnarray}
\label{SCETfielddef}
\chi_n = W_n \xi_n\,, \qquad {\cal A}_n^\mu = \frac{1}{\bnp} W_n[\nbar\cdot D,D^\mu] W_n^\dagger\,. \label{cadef}
\end{eqnarray}
In the SCET literature ${\cal A}_\mu$ is commonly written as $B^\perp_\mu$. 

To reproduce the production of two collinear back-to-back partons one requires an operator in SCET which contains 
two collinear fields, one for each of the directions:
\begin{equation}
\label{O2def}
\cO_2^{(n,\nbar)} = \bar \chi_n \Gamma \chi_\bn\,.
\end{equation}
Each operator comes with a set of labels, 
corresponding to the directions $n_i$ of its collinear fields. We write
\begin{equation}
\cO_j^{(n)} \sim \cO_2^{(n_1, n_2)}, \cO_3^{(n_1, n_2, n_3)}, \cdots\,,
\end{equation}
where $n_j$ are the label momenta. 
The labels identify the degrees
of freedom which cannot change -- all the hard degrees of freedom which could have changed the collinear
momentum in the full theory are integrated out of SCET. Note that there can be many operators with
the same labels but different tensor structure (for example, the operators $\cO_3$ and $\cO_3^{(2)}$
we will define later on). 
For notational simplicity, we will often omit the labels $(n_i)$ when there
is no ambiguity.

The matching between QCD and SCET implies that we want to choose
coefficients $\cC_j$ for these operators such that the sum over all the operators 
reproduces the current of QCD
\begin{equation}
\cJ = \cC_2 \cO_2 + \cC_3 \cO_3+\cC_4 \cO_4 +\cdots \,.
\end{equation}
Since the Wilson coefficients $\cC_j$ only encode short distance physics,
they are independent of the choice of states used in the calculation of matrix elements.
So, to satisfy this equation, 
we can take matrix elements in convenient states, and build up the $\cC_j$ systematically.
First, we take matrix elements with two quarks in the final state, which gives the matching condition
\begin{equation}
\langle \cJ|q \qbar\rangle = \cC_2\langle \cO_2 | q\qbar\rangle\,,
\end{equation}
as no other higher order operators have matrix elements in a 2-quark state. This allows us to 
determine $\cC_2$. Next, we take matrix elements with two quarks and one gluon
\begin{equation}
\langle \cJ|q \qbar g\rangle = \cC_2  \langle \cO_2| q\qbar g\rangle+ \cC_3 \langle \cO_3| q \qbar g \rangle\,,
 \label{mes}
\end{equation}
and allows us to determine $\cC_3$. The Wilson coefficients $\cC_j$ with $j>3$ are determined analogously. 
In order to correctly match QCD with up to $m$ well separated partons requires operators $\cO_j$ with 
$n \le m$ in SCET. 

The next step is running.
The matching determines the Wilson coefficients at the matching scale, $\cC_j(\mu=Q)$ and the running
will allow us to obtain them at a lower scales. The Wilson coefficients at these two scales are related by
\begin{eqnarray}
\cC_n(\mu) = \cC_n(Q) \Pi_n(Q,\mu)\,.
\end{eqnarray}
The calculation of  the evolution kernel $\Pi_n$
is a straightforward application of the renormalization group, and involves
calculating the anomalous dimensions of the operators in SCET. Because the interactions in SCET are
simpler than in QCD, the anomalous dimensions are fairly easy to compute. In fact,
we will compute a closed form, algebraic expression, for the LL anomalous dimension of any $\cO_j$.

The interactions in SCET allow for collinear gluons to be radiated off collinear quarks and gluons, 
or collinear gluons to split into two collinear quarks. However, since SCET only describes the 
collinear or soft limit of QCD, the transverse momentum between the resulting particles has to be small. 
How small depends on the renormalization scale at which the emission is calculated, and 
the requirement is typically $p_\perp \lesssim \mu$. 
As an example consider the matrix element of $\cO_2$ in an three parton final state with 
specific momenta $|q\qbar g\rangle$.
Let $p_T$ be the transverse momentum of the gluon with respect to a quark.\footnote{
Our notation is that 
$p_\perp$ is SCET notation for the perpendicular momentum label on a collinear field, while $p_T$ refers to
the transverse momentum between two four-vectors.}
If the renormalization scale satisfies $\mu > p_T$, then gluon emission in SCET can give rise 
to a gluon in the final state, and a non-vanishing matrix element. If, however, $\mu < p_T$, then 
the SCET emission is not able to produce a gluon in the final state, and the matrix element of $\cO_2$ 
vanishes. 

For theory to be continuous across $p_T$, 
matrix elements at the scale $\mu = p_T-\epsilon$ must be equal to matrix elements at 
$\mu = p_T + \epsilon$. 
More precisely, we require that
\begin{equation}
[\cC_2 \langle \cO_2 \rangle + \cC_3 \langle \cO_3 \rangle +\cdots ]_{\mu = p_T + \epsilon} = 
[\cC_2 \langle \cO_2 \rangle + \cC_3 \langle \cO_3 \rangle +\cdots ]_{\mu = p_T - \epsilon}
\end{equation}
To compensate for the discontinuity of $\langle \cO_2 \rangle$ at $\mu=p_T$, we only need to
change $\cC_3 \langle \cO_3 \rangle$;  all other matrix elements are continuous across the threshold.
So we derive the matching condition
\begin{equation}
\label{thresholdmatching}
[\cC_2 \langle \cO_2 \rangle ]_{\mu = p_T+\epsilon}=
[\cC_3 \langle\cO_3^{(2)} \rangle ]_{\mu = p_T-\epsilon}\,.
\end{equation}
We call this matching threshold matching and it
is done wholly within SCET. 
Note that the operator $\cO_3^{(2)}$ in Eq.~(\ref{thresholdmatching}) is different from the operator $\cO_3$ arising in the hard matching at $\mu =Q$. To distinguish these operators we add a superscript, which labels the number of partons present at the hard matching scale. For simplicity, we omit the superscript for the operators
matched at $Q$: $\cO_j \equiv \cO_j^{(j)}$. 

Let us apply these results and calculate the SCET expressions for a three jet final state, with 
the transverse momentum of the emitted gluon given by $p_T$. After matching QCD onto SCET at the hard scale $\mu = Q$ we find
\begin{eqnarray}
\label{matchschematic}
\langle {\rm SCET} \rangle_Q = \cC_2(Q) \langle \cO_2 \rangle + \cC_3(Q) \langle \cO_3 \rangle 
= \langle {\mathrm {QCD}} \rangle
\end{eqnarray}
Using the RG evolution, we can obtain this at a lower scale $\mu$
\begin{eqnarray}
\langle {\rm SCET} \rangle_\mu  = \cC_2(Q) \Pi_2(Q,\mu) \langle \cO_2 \rangle + \cC_3(Q) \Pi_3(Q,\mu)
\langle \cO_3 \rangle\,.
\end{eqnarray}
Finally, at the $\mu = p_T$ we have to perform the threshold matching to obtain
\begin{eqnarray}
\label{SCET_pT1}
\langle {\rm SCET} \rangle_{p_T}= \cC_2(Q) \Pi_2(Q,p_T) \langle \cO_3^{(2)} \rangle
+ \cC_3(Q) \Pi_3(Q,p_T)\langle \cO_3 \rangle\,.
\end{eqnarray}

To calculate a differential cross section, we need to square this matrix element and sum over final state spins and polarizations. Using the Feynman rules in SCET we will show later that 
emissions in SCET factorize. In particular, 
\begin{eqnarray}
\label{O2simplify}
\left| \langle \cO_3^{(2)}| q \qbar g \rangle \right|^2 \approx  \left| \langle \cO_2| q \qbar \rangle \right|^2 P(p_T,z)
\end{eqnarray}
where $z = E_q/(E_g+E_q)$.
The function $P(p_T,z)$ is equivalent to a splitting function of QCD, up to power corrections, in 
the collinear limit.

We can now show that SCET agrees with QCD for large $p_T$ and the parton shower for small $p_T$.
First, consider the limit $p_T \sim Q$. Then,
$\Pi_n(Q,p_T) \sim 1$ up to higher order corrections. 
So Eq.~\eqref{SCET_pT1} reduces to Eq.~\eqref{matchschematic} and we reproduce QCD.
Next, take the limit $p_T \ll Q$. 
Then we are in the collinear limit, and so the matrix element 
of $\cO_2$ is very similar to that of QCD. Thus,
$\langle \cO_3\rangle\approx 0$, since $\cO_3$ is the 
difference between QCD and SCET. So 
$\rd \sigma \sim \cC_2(Q)\Pi_2^2 \langle \cO_3^{(2)}\rangle^2$. 
Now, the kernel $\Pi_2$, is equivalent to the Sudakov factor, 
and as shown in Eq.~(\ref{O2simplify}), the matrix element reproduces 
a splitting function,
therefore the SCET differential cross section in this limit reduces to the parton shower result~\eqref{dsps}.

\section{Required calculations in SCET}
\label{sec:details}
In this section we present details of the SCET calculations required to implement the
scheme from the previous section. As we have seen in the previous section, 
there are three steps in the SCET calculations required to obtain jet distributions. 
First, we need to calculate the matching from QCD to SCET at some hard scale $\mu \sim Q$. 
Second, the renormalization scale of the operators is 
lowered using the renormalization group evolution. 
Third, a threshold matching is required when the renormalization scale 
gets lowered past the transverse momentum of one of the partons in the final state. 
Each of these three steps will be addressed
in its own subsection. We will try to keep results as general as possible,
but sometimes it will be necessary to choose a particular example. 
In this section we assume that the reader is familiar with the basic idea of SCET. 
For a quick review of SCET we refer the reader to 
Appendix~\ref{app:scet} and the original literature~\cite{SCET1,SCET2,SCET3,SCET4}. 
This is the most technical section
of the paper, and for readers not interested in the details, we will briefly summarize 
the results obtained in this section. This will make it possible to skip this section and still be 
able to follow the rest of the paper.

The SCET results at  ${\mathcal O}(\alpha_s)$ are as follows.
The full QCD current is reproduced in the effective theory by operators $\cO_n$, such that 
\begin{equation}
 \cJ= \cC_2 \cO_2+ \cC_3 \cO_3 + \cC_3^{(2)}  \cO_3^{(2)} + \cdots \,,
\end{equation}
The operators are given by
\begin{eqnarray}
\cO_2 &=& \bar \chi_n \Gamma \chi_{\bn} \\
\cO_3 &=& g_s \overline{\chi}_{n_q}\left[
\Slash{\cal A}_{n_g}\frac{\nsl_\qbar}{2} \frac{1}{ n_\qbar \tdot \cP^\dagger} \Gamma- \Gamma 
\frac{1}{ n_q \tdot \cP}
\frac{\nsl_q}{2}
 \Slash{\cal A}_{n_g}\right]\chi_{n_\qbar} \\
\cO_3^{(2)} &=& g_s 
\overline{\chi}_{n_q}\left[
\cAsl_{n_g} 
\frac{\bnslash_{\bar q}}{2}  \frac{1}{\bn_{\bar q} \tdot {\cal P}^\dagger}\Gamma
-\Gamma
\frac{1}{\bn_q \tdot \cal P} \frac{\bnslash_q}{2}
\cAsl_{n_g} 
\right]
\chi_{n_\qbar}\, ,
\label{o3op}
\end{eqnarray}
and for the Wilson coefficients required to NLO we find
\begin{eqnarray}
\cC_2(Q) &=& 1 - \frac{\alpha_s C_F}{4 \pi} \left(8 - \frac{7 \pi^2}{6}  +3 \pi i \right)\\
\cC_3(Q) &=& 1\\
\cC_3^{(2)}(p_T) &=& \cC_2(p_T)
\,.
\end{eqnarray}
The Wilson coefficients $\cC_n$ satisfy the RG equation
\begin{eqnarray}
\mu \frac{d}{d\mu}\cC_n(\mu)  = \gamma_n (\mu) \cC_n(\mu)
\end{eqnarray}
The anomalous dimensions are
\begin{eqnarray}
\gamma_2
&=& -\frac{\alpha_s}{\pi} \left[{\ccol C_F} \log\frac{-\mu^2}{\QQ} +\dcol C_F\frac{3}{2}\right]
\label{gamma2}
\\
\gamma_3 &=& 
 -\frac{\alpha_s}{ \pi} 
\left[C_F \left( \log\frac{-\mu^2}{(\pq+ \pqb)^2} + \frac{3}{2} \right)
+ \frac{C_A}{2} \left( \log \frac{-\mu^2 (\pq+ \pqb)^2}{(\pq + \pg)^2(\pqb + \pg)^2} + \frac{11}{6} \right)
- \frac{n_f}{6}
\right]\,,
\label{gamma3}
\\
\gamma_n &=& -\frac{\alpha_s}{ \pi} \left[{\ccol 
\left( \frac{n_q}{2} C_F + \frac{n_g}{2} C_A\right)} \log\frac{\mu^2}{Q^2} +
\dusp_n \right] 
\label{gamman}
\end{eqnarray}
The Wilson coefficients evolve through the kernels $\Pi_n$:
\begin{eqnarray}
\cC_n(\mu_1) = \cC_n(\mu_2) \Pi_n(\mu_2,\mu_1)
\end{eqnarray}
For a general anomalous dimension
\begin{equation}
\gamma_n 
= 
-\frac{\alpha_s}{\pi}\left(\cusp_n \log\frac{\mu^2}{Q^2} + \dusp_n 
\right)
\end{equation}
we can solve $\Pi_n$ explicitly
\begin{eqnarray}
\Pi_n(Q,\mu) &=& \exp \left\{
\frac{8\pi}{\beta_0^2 \alpha_s(Q)}
\cusp_n\left(\log\frac{\alpha_s(Q)}{\alpha_s(\mu)}+1-\frac{\alpha_s(Q)}{\alpha_s(\mu)}\right)
-\frac{2}{\beta_0} \dusp_n \log\frac{\alpha_s(Q)}{\alpha_s(\mu)}\right\}
 \nn\\
&=& 1-\frac{\alpha_s(Q)}{4 \pi} \left[ 
\cusp_n \log^2\frac{\mu^2}{Q^2}
+2 \dusp_n\log\frac{\mu^2}{Q^2}
\right] 
+\cdots \,, \label{pin}
\end{eqnarray}
where
\begin{eqnarray}
\alpha_s(\mu) = \frac{\alpha_s(Q)}{1 + \frac{\beta_0}{4 \pi} \alpha_s(Q) 
\log\frac{\mu^2}{Q^2}}\,,\quad \beta_0 = \frac{11}{3}C_A - \frac{2n_f}{3}
\end{eqnarray}
Only the piece proportional to $\cusp_n$ is the leading log resummation, the $\dusp_n$ resums 
a subset of next-to-leading logs.

\subsection{Matching from QCD to SCET}
The first step is matching from QCD to SCET. 
As discussed in Section \ref{sec:schematic}, the matching condition is 
\begin{equation}
 \cJ= \cC_2 \cO_2+ \cC_3  \cO_3+\cC_4  \cO_4 +\cdots \label{ope2}\,,
\end{equation}
and we want to choose Wilson coefficients
in the effective theory such that QCD is reproduced 
at the scale $Q$ to a given order in perturbation theory. QCD matrix elements
with up to $n$ well separated partons appear at ${\cal O}(g_s^{n-2})$ in perturbation theory. 
In order to correctly reproduce these matrix elements, we require operators in SCET with up to $n$ 
collinear fields. In the remainder of this section, we will explicitly perform the matching 
onto operators $\cO_2$ and $\cO_3$, and comment on how to extend these calculations to 
operators with four or more collinear fields. 

As discussed above, each operator $\cO_j$ depends on $j$ labels for the 
directions of the $j$ collinear fields. 
For each set of labels a different Wilson coefficient exists, so that in principle each 
product of operator and Wilson coefficient in Eq.~(\ref{ope2})
 represents an infinite sum over the 
various labels. We will carefully treat the label dependence in the two-jet matching, but will then neglect the label dependence in the further discussions. 

\subsubsection{Matching onto $\cO_2$ at tree level}
The first step is to ensure that matrix elements with two quarks are correctly reproduced. 
Since the operators $\cO_j$ with $j>2$ have at least 3 collinear fields, they do not contribute to these
matrix elements. Thus we need to choose Wilson coefficients $\cC_2^{(n)}$ so that
\begin{equation}
\langle \cJ | q \qbar\rangle = \sum_{n_i}\cC_2^{n_1 n_2} \langle \cO^{(n_1,n_2)}_2| q \qbar\rangle\,,
\label{jet2match} 
\end{equation}
where
\begin{equation}
 \cO^{(n_1,n_2)}_2 =  \bar{\chi}_{n_1} \Gamma \chi_{n_2} 
\end{equation}
is a general basis of operators with 2 collinear fermion fields.
To evaluate matrix elements on the right hand side of \eqref{jet2match}, we need to know how jets $\chi_n$ 
act on quark states $|q\rangle$. The simplest
prescription is
\begin{equation}
 \chi_n | q\rangle = \delta_{n^\mu, n^\mu_q}\,, \label{nqpre}
\end{equation}
where $n^\mu_q = p^\mu_q/E_q$. 
If $\cJ$  produces two quarks, they must be back-to-back in the center-of-mass frame. 
So with this prescription,
only the operators with $n_2 = \bn_1$ get a non-zero coefficient. With these conventions, the  matrix elements of the operator $\cO_2$ is identical to the matrix element of the QCD current $\cJ$ and 
we thus obtain
\begin{equation}
\cC_2^{n, \nbar} = 1 \,, \quad\quad
\cC_2^{n_1, n_2} = 0 \,,\,\, n_1 \ne \nbar_2\,.
 \label{c2nn} 
\end{equation}
SCET fields also have labels corresponding to their $p_\perp$ momenta, but we choose to turn off any operator
with $p_\perp \ne 0$. 

Other choices for collinear fields acting on quark states are possible, and in fact \eqref{nqpre} is by itself ambiguous.
We can always write $\chi_n$ as a jet in a different direction $n'$ using \eqref{diractojet}
\begin{eqnarray}
\label{n1n2relation}
\chi_{n} = \frac{\nslash \bnslash}{4} 
\left[1 + \frac{\pperpsl}{\bn' \tdot p} 
\frac{\bnslash'}{2} \right] \chi_{n'}\,, 
\end{eqnarray}
where $p_{\perp}$ is transverse to $n'$. Thus it would seem that
$\chi_n | q\rangle\ne0$ even if $q$ is not aligned with $n$. 
This ambiguity is resolved at higher order. 
Note that $n \tdot n' \sim p_{\perp}^2$, thus as 
long as $n\tdot n' \ll 1$ the two fields are equivalent up 
to power corrections. This implies that any operator 
$\bar \chi_{n_1} \Gamma \chi_{n_2}$ with $n_1 \tdot n_2 \ll 1$ 
could be used as the operator $\cO_2$. 
Of course, in order to determine the matrix element of the 
operator $\cO_2$ requires knowledge of how a SCET field with a 
given direction $n$ creates or annihilates a final state with 
momentum in a direction $n'$. Any choice other than Eq.~\eqref{nqpre} will make 
the required calculations more difficult, so for simplicity we will stick to \eqref{nqpre}
for the rest of this paper.

\subsubsection{Matching onto $\cO_3$ at tree level}
Now that we have determined the matching onto the operator $\cO_2$ we can 
proceed to match with 3-parton final states. The matching condition reads
\begin{equation}
\langle \cJ| q \qbar g \rangle= 
\cC_2 \langle \cO_2|q \qbar g \rangle+
\cC_3 \langle \cO_3|q \qbar g \rangle\,. \label{match3}
\end{equation}
The left hand side is given by the matrix elements for real emission in QCD
from the quark and the antiquark leg
\begin{eqnarray}
\label{QCDme}
\langle\cJ |q \qbar g\rangle^{q}&=&g_s \overline{\psi}_q\frac{\Asl (\qsl + \gsl)}{(\pq + \pg)^2} \Gamma \psi_\qbar 
\nn \\
\langle\cJ |q \qbar g\rangle^{\qbar} &=& -g_s
 \overline{\psi}_q\Gamma\frac{(\qbsl + \gsl)\Asl}{(\pqb + \pg)^2} \psi_\qbar\,.
\end{eqnarray}

For the right hand side, the operator $\cO_2$ and its Wilson coefficient were determined in the previous section, but 
we need to evaluate the three-parton matrix element for $\cO_2$. In principle, the additional gluon can 
be either collinear or soft. We will first choose the gluon to be collinear and later check that the 
resulting matching condition is satisfied in the soft limit as well. The
collinear emissions in SCET can come out of either a vertex from the Lagrangian or from 
a Wilson line. The matrix elements
are extracted from the interaction vertices, and simplify with the equations of motion 
$\nbsl \xi_\nbar = \overline{\xi}_n \nsl = 0$.
\begin{eqnarray}
\langle \cO_2| q \qbar g \rangle^{L_q} &=& 
g_s
\frac{\nbar \tdot (\pq + \pg)}{(\pq+\pg)^2} 
\overline{\xi}_n 
A_\mu
\left[
n^\mu 
+\frac{\qsl^\perp \gamma^\mu_\perp}{\nbar \tdot \pq} 
+\frac{\gamma^\mu_\perp (\qsl^\perp +\gsl^\perp)}{\nbar \tdot( \pq +\pg)}
- \frac{\qsl^\perp(\qsl^\perp+\gsl^\perp)}
{\nbar \tdot \pq \nbar \tdot (\pq+\pg)}\nbar^\mu
\right]  
\Gamma
\xi_\nbar
\nn\\
\langle \cO_2| q \qbar g \rangle^{L_\qbar} &=& 
-g_s\frac{n \tdot (\pqb + \pg)}{(\pqb+\pg)^2} 
\overline{\xi}_n 
\Gamma  
\left[\bn^\mu
+\frac{\gamma^\mu_\perp \qbsl^\perp}{n \tdot \pqb} 
+\frac{(\qbsl^\perp+\gsl^\perp) \gamma^\mu_\perp}{n \tdot( \pqb +\pg)}
- \frac{(\qbsl^\perp+\gsl^\perp) \qbsl^\perp}
{n \tdot (\pqb+\pg)n \tdot \pqb} n^\mu
\right] 
A_\mu
\xi_\nbar\nn
\end{eqnarray}
\begin{equation}
\langle \cO_2| q \qbar g \rangle^{W_q}
 =  g_s  \overline{\xi}_n  \frac{\nbar \tdot A}{\bar n\tdot\pg}\Gamma \xi_\nbar, 
\qquad
\langle \cO_2| q \qbar g \rangle^{W_\qbar} 
= -g_s \overline{\xi}_n \Gamma  \frac{n\tdot A}{n\tdot\pg} 
\xi_\nbar
\end{equation}

As the SCET Lagrangian is constructed to be gauge invariant, the matrix elements should satisfy a
Ward identity. It is a straightforward check to see that $\langle \cO_2| q \qbar g \rangle^{W}
+ \langle \cO_2| q \qbar g \rangle^{L}$ vanishes when 
$A^\mu=\pg^\mu$. This holds separately for the quark and antiquark emissions, as it must since SCET
is invariant under gauge transformations of the collinear fields in each direction separately. 

There are several issues that have to be resolved before we can use the above results 
to determine the matching of the operator $\cO_3$. First, note that the collinear 
emission from a Wilson line 
gives a divergence if the large label momentum ($\bn \tdot p_g$ for $\langle \cO_2 \rangle^{W_q}$ 
and $n \tdot p_g$ for $\langle \cO_2\rangle^{W_{\bar q}}$) becomes small. 
In SCET, the collinear phase space integration involves summing over the large momentum labels, 
as well as integrating over the residual part of the momentum. However, in the sum
over collinear momenta the value $\bn \cdot p = 0$ and $p^\perp=0$ has to be omitted~\cite{zerobin}. 
Thus, the phase space integration will never reach  a value for the label momentum such that
the unphysical divergence is realized. However, keeping track of the condition  $\bn \cdot p \neq 0$ 
in the 
phase space integration can be difficult, and we therefore propose a scheme which will allow us 
to integrate the phase space naively, without having to worry about small label momenta. 
In this scheme one neglects the 
emissions from the collinear Wilson lines all together, and ensures gauge invariance by summing
over only transverse polarizations when squaring the resulting matrix elements. 
By neglecting the $\langle \cO_2 \rangle^W$ contributions, we subtract the divergences at small 
$\bn \tdot p$.
Any appropriate prescription
will modify the effective theory only at short distances, 
and therefore differences in prescription can be absorbed by appropriately adjusting 
the Wilson coefficients in the matching condition. 
In practice, however, we have to evaluate matrix elements at finite momenta, which are 
neither exactly at the hard scale
where the matching is done, nor in the strict collinear limit, 
where the different prescriptions coincide.
Therefore the physical results may differ. 
However, these differences are beyond leading order in the SCET
expansion, {\it i.e.} they are power corrections.

The second issue is that emissions in SCET cannot change the directions of the fermions. 
Since the operators 
$\cO_2$ that were matched on in the two jet matching have the two fermions 
in back-to-back directions, the matrix element of $\cO_2$ can only give rise to final 
state with back-to-back fermions. But if momentum is to
be conserved when we include the gluon, the directions of the quark 
and anti-quark momenta can never satisfy $n_q = \overline{n_\qbar} $ exactly. 
This implies that in order to insist on \eqref{nqpre}, we need to 
change the direction of at least one of $\bar{\xi}_n$ or $\xi_\nbar$ using \eqref{n1n2relation},
otherwise we could never get a non-vanishing matrix element with three partons in different 
directions.
There are again various possibilities to deal with this, and all of them will lead to the 
same results up to power corrections. The choice we will adopt here is that directions
of collinear fields which are not involved in the emission do not change. Then we rotate the
emitting field into the direction of its final state. In other words, we use
\begin{equation}
\overline{\xi}_n \cX [\Gamma \xi_\nbar] \to  \overline{\xi}_{n_q} \frac{\nsl_\qbar \nbsl_\qbar}{4}  
\cX [\Gamma \xi_{n_\qbar}], 
\qquad
[\overline{\xi}_n \Gamma]  \cX \xi_\nbar \to [\overline{\xi}_{n_q}\Gamma]  \cX \frac{\nbsl_q\nsl_q}{4}  \xi_{n_\qbar}\,,
\label{jetproj}
\end{equation}
where $\cX$ is some arbitrary operator. This equation shows how to replace a fermion in SCET with a fermion in QCD, which
is necessary to compare matrix elements in the same external states. 

We now see that two {\it different} operators
$\cO_2^{(n_\qbar \nbar_\qbar)}$ and  $\cO_2^{(n_q \nbar_q)}$ 
contribute to the same final state. Pictorially,
\begin{equation}
\SetScale{0.7}
  \begin{picture}(100,60)(60,0)
 \SetWidth{0.5}
    \SetColor{Black}
    \COval(57,56)(7.07,7.07)(135.0){Black}{White}\Line(53.46,59.54)(60.54,52.46)\Line(60.54,59.54)(53.46,52.46)
    \SetWidth{1.0}
    \LongArrow(71.22,43.45)(107.22,25.45)\LongArrow(70.78,42.55)(106.78,24.55)
    \LongArrow(257.22,39.55)(226.22,24.55)\LongArrow(256.78,40.45)(225.78,25.45)
    \SetWidth{0.5}
    \COval(170,12)(7.07,7.07)(135.0){Black}{White}\Line(166.46,15.54)(173.54,8.46)\Line(173.54,15.54)(166.46,8.46)
    \COval(279,50)(7.07,7.07)(135.0){Black}{White}\Line(275.46,53.54)(282.54,46.46)\Line(282.54,53.54)(275.46,46.46)
    \ArrowLine(273,54)(246,74)
    \ArrowLine(49,56)(14,56)
    \ArrowLine(64,56)(99,56)
    \ArrowLine(285,45)(312,25)
    \Gluon(176,16)(202,50){5}{3.58}
    \ArrowLine(176,9)(203,-11)
    \ArrowLine(163,12)(128,12)
  \end{picture}\nn
\end{equation}
The first will turns $\cO_2$ into $\cO_3^{(3)}$ when the quark emits, and the second
when the antiquark emits. We want to emphasize again that this is just a convention,
and equivalent to any other convention up to power corrections.

The transverse momenta are measured with respect to the directions $n$ and $\bn$. This implies 
that when the quark emits, $\pq^\perp +\pg^\perp=0$ and when the antiquark emits,
$\pqb^\perp +\pg^\perp=0$. Using these results together with \eqref{jetproj}, the quark emission
and antiquark emission contributions to the matrix element of $\cO_2$ can be simplified to
\begin{eqnarray}
\label{O2me}
\langle \cO_2| q \qbar g \rangle^{L_q}&=& 
g_s \frac{n_\qbar \tdot( \pq + \pg)}{(\pq+\pg)^2} 
\overline{\xi}_{n_q}\Asl \frac{\nbsl_\qbar}{2} \Gamma \xi_{n_\qbar}
\nn \\
\langle \cO_2| q \qbar g \rangle^{L_{\bar q}} &=& 
-g_s \frac{n_q \tdot( \pqb + \pg)}{(\pqb+\pg)^2} 
\overline{\xi}_{n_q}\Gamma \frac{\nbsl_q}{2} \Asl \xi_{n_\qbar}
\label{mconv}
\end{eqnarray}

Using the results of the SCET matrix elements given in Eq.~\eqref{O2me} and
 the QCD matrix elements given in Eq.~\eqref{QCDme}, their difference reduces to the simple form\cite{BS} 
\begin{eqnarray}
\langle\cJ |q \qbar g\rangle^{q} - \langle \cO_2| q \qbar g \rangle^{L_q} &=&
g_s \frac{\nbar_\qbar \tdot( \pq + \pg)}{(\pq+\pg)^2} 
\overline{\xi}_{n_q}\Asl \frac{\nsl_\qbar}{2} \Gamma \xi_{n_\qbar}
\nn \\
\langle\cJ |q \qbar g\rangle^{\qbar} - \langle \cO_2| q \qbar g \rangle^{L_\qbar} 
&=& 
-g_s \frac{\nbar_q \tdot( \pqb + \pg)}{(\pqb+\pg)^2} 
\overline{\xi}_{n_q}\Gamma \frac{\nsl_q}{2} \Asl \xi_{n_\qbar}
\label{mdiff}
\end{eqnarray}

The final step in the matching is to choose a basis for the 3-jet operators $\cO_3^{(n)}$. 
The standard
convention is to have $\cC_3=1$ at tree level, so we take $\cO_3 = \cJ- \cO_2$. 
Thus,  we arrive at
\begin{equation}
\cO_3 = g_s \overline{\chi}_{n_q}(
\Slash{\cal A}_{n_g}\frac{\nsl_\qbar}{2} \frac{1}{ n_\qbar \tdot \cP^\dagger} \Gamma- \Gamma 
\frac{1}{ n_q \tdot \cP}
\frac{\nsl_q}{2}
 \Slash{\cal A}_{n_g}
)\chi_{n_\qbar}, \qquad
\cC_3 =1\,,
\end{equation}
where ${\cal A}_n^\mu$ was defined in Eq.~\eqref{SCETfielddef}. 
Note that $\cO_3$ is the difference between the QCD and SCET emissions. 
Since the operator $\cO_2$, together with the SCET emission of an additional gluon  
describes the infrared physics of QCD, the operator $\cO_3$ only has contributions for 
large values of the transverse momentum. 

If the gluon is soft instead of collinear, there are two emission diagrams in SCET. 
The matrix elements are
\begin{equation}
\langle \cO_2| q \qbar g \rangle^{S_q} 
=g_s \overline{\xi}_n  \frac{n\tdot A}{n\tdot\pg}\Gamma \xi_\nbar\,, \qquad
\langle \cO_2| q \qbar g \rangle^{S_\qbar} 
=-g_s \overline{\xi}_n \Gamma\frac{\nbar \tdot A}{\bar n\tdot\pg} \xi_\nbar\,.
\end{equation}
This can be compared with the collinear results given our convention of dealing with 
label momenta going to zero. 
In that scheme we only keep the matrix elements $\langle \cO\rangle ^{L}$
 and sum only over transverse polarizations of the gluons. 
Taking the soft limit of $\langle \cO_2 | q \bar q g \rangle$ 
(by taking $p_g \ll \bn \tdot p_q,n \tdot p_{\bar q}$) we find
\begin{equation}
\label{O2soft}
\langle \cO_2| q \qbar g \rangle^{L_q} 
\to g_s \overline{\xi}_n  \frac{n\tdot A}{n\tdot\pg}\Gamma \xi_\nbar\,, \qquad
\langle \cO_2| q \qbar g \rangle^{L_\qbar} 
\to - g_s \overline{\xi}_n \Gamma\frac{\nbar \tdot A}{\bar n\tdot\pg} \xi_\nbar\,.
\end{equation}
Thus, the soft limit of the collinear Lagrangian emission is identical to the soft gluon emission. 
However, care has to be taken about the second part of the convention, namely the fact that we only sum 
over transverse polarizations. Since for gauge invariant amplitudes the longitudinal 
polarizations do not contribute to physical processes, the soft gauge invariance of SCET
is enough to ensure that the sum over transverse polarizations is equivalent to the sum 
over all polarizations. SCET is by construction gauge invariant, but one can also 
see the invariance trivially from Eq.~\eqref{O2soft}. Replacing $A^\mu \to p_g^\mu$, the 
two contributions are equal and opposite in sign and thus cancel. Thus, the amplitude satisfies
the Ward identity and is gauge invariant. With these two observations it is obvious that the 
collinear emission reproduces the soft emission properly. 

In summary, we have chosen the conventions
\begin{itemize}
\item All fields appearing in operators have labels for collinear momenta only; in our basis,
 all $p_\perp$ momentum components are zero.
\item Emissions are calculated with the collinear emission from the SCET Lagrangian only. 
Wilson line emission and soft emission are discarded. 
\item To maintain gauge invariance, we include only transverse polarizations of the gluons.
\item Fermions in SCET are rotated to a direction aligned with their 4-momentum before matrix elements are taken. 
{\it I.e.} $\xi_n \to \frac{\nsl \nbsl}{4} \xi_{n_q}$, where $n^\mu_q =\frac{1}{E_q} q^\mu$ for massless fields. 
\end{itemize}
All other conventions are equivalent up to power corrections.

\subsubsection{Matching onto $\cO_4$ at tree level}
We can outline the calculation for matching
4-jet operators $\cO_4$. These will make up for the difference between
what SCET predicts when we have used only 3-parton QCD results, and the true 4-parton QCD prediction. 
Starting from $q\qbar$, a 4-parton state can be either $q g g \qbar$ or $q \qbar q \qbar$. We need to solve
\begin{equation}
\langle \cJ| 4 \rangle= \sum_{(n)}
\cC_2^{(n)} \langle \cO_2^{(n)}|4 \rangle+
\cC_3^{(n)} \langle \cO_3^{(n)}|4 \rangle+
\cC_4^{(n)} \langle \cO_4^{(n)}|4 \rangle \label{match4}
\end{equation}
Let us take the $|q \qbar q \qbar\rangle$ state as an example. Then there are two QCD diagrams which 
contribute
\begin{equation}
\langle \cJ | \qqqq \rangle =
\SetScale{0.5}
  \begin{picture}(80,35) (0,10)    
    \SetWidth{0.5}
    \SetColor{Black}
    \ArrowLine(60,35)(0,35)
    \ArrowLine(60,35)(90,35)
    \Gluon(90,35)(116,56){5}{3}
    \ArrowLine(120,95)(115,56)
    \ArrowLine(150,80)(116,57)
    \Photon(50,35)(50,0){4}{3}
    \ArrowLine(90,35)(140,35)
  \end{picture}
+
\SetScale{0.5}
  \begin{picture}(80,35) (-5,10)    
    \SetWidth{0.5}
    \ArrowLine(50,34)(0,34)
    \ArrowLine(80,34)(140,34)
    \ArrowLine(80,34)(50,34)
    \Photon(80,34)(80,0){4}{3}
    \Gluon(50,34)(70,54){5}{3}
    \ArrowLine(104,78)(70,54)
    \ArrowLine(70,54)(81,94)
   \end{picture}
\end{equation}
On the SCET side, the $\cO_2$'s can contribute
\begin{equation}
\langle \cO_2 | \qqqq \rangle =
\SetScale{0.5}
  \begin{picture}(90,25) (0,5)    
    \SetWidth{0.5}
    \SetColor{Black}
    \ArrowLine(60,15)(0,15)
    \ArrowLine(60,15)(90,15)
    \ArrowLine(90,15)(135,0)
    \Gluon(90,15)(116,36){5}{2}
    \ArrowLine(120,75)(115,36)
    \ArrowLine(150,60)(116,37)
    \COval(60,15)(7.07,7.07)(135.0){Black}{White}\Line(56.46,18.54)(63.54,11.46)\Line(63.54,18.54)(56.46,11.46)
    \end{picture}
+
\SetScale{0.5}
  \begin{picture}(90,25) (0,10)    
    \SetWidth{0.5}
    \ArrowLine(77,86)(64,42)
    \ArrowLine(105,70)(64,42)
    \Gluon(45,23)(64,42){5}{2}
    \ArrowLine(45,23)(10,23)
    \ArrowLine(72,13)(45,23)
    \ArrowLine(80,11)(115,-1)
    \COval(77,12)(7.07,7.07)(135.0){Black}{White}\Line(73.46,15.54)(80.54,8.46)\Line(80.54,15.54)(73.46,8.46)
    \end{picture}
\end{equation}
as well as $\cO_3$
\begin{equation}
\langle \cO_3 | \qqqq \rangle =
\SetScale{0.5}
  \begin{picture}(90,25) (0,5)    
    \SetWidth{0.5}
    \SetColor{Black}
    \ArrowLine(54,4)(89,-8)
    \ArrowLine(85,70)(72,26)
    \ArrowLine(112,55)(72,26)
    \ArrowLine(41,6)(6,6)
    \Gluon(52,9)(72,26){5}{2}
    \COval(48,6)(7.07,7.07)(135.0){Black}{White}\Line(44.46,9.54)(51.54,2.46)\Line(51.54,9.54)(44.46,2.46)
    \end{picture}
\end{equation}
and of course $\cO_4$:
\begin{equation}
\langle \cO_4 | \qqqq \rangle =
\SetScale{0.5}
  \begin{picture}(90,25) (0,5)    
    \SetWidth{0.5}
    \SetColor{Black}  
    \ArrowLine(45,12)(10,12)
    \ArrowLine(62,60)(49,16)
    \ArrowLine(93,41)(53,12)
    \ArrowLine(54,10)(89,-2)
    \COval(50,12)(7.07,7.07)(135.0){Black}{White}\Line(46.46,15.54)(53.54,8.46)\Line(53.54,15.54)(46.46,8.46)
    \end{picture}
\end{equation}
Using the results for $\cO_2$ and $\cO_3$ obtained previously, 
we can determine $\cC_4$ and $\cO_4$. In a similar way, the tree-level matching for $n$ jets can be worked out.
Note that since we have already set the conventions for evaluating matrix elements, there is no additional ambiguity when
we match to operators with 4 or more collinear fields.

\subsubsection{2-jet matching at NLO \label{subsubsec:2jetNLO}}
So far, we have performed the matching calculations at tree level, and the normalization of the 
operators was chosen such that all Wilson coefficients are unity. In this subsection, we will 
determine the Wilson coefficient $\cC_2$ at ${\cal O}(\alpha_s)$, which will be required to describe
jet distributions at next-to-leading order (NLO). The matching condition is still 
given by Eq.~\eqref{jet2match}, but the matrix elements need now be evaluated at one loop. 
Loop diagrams are in general both IR and UV divergent, and we regulate IR divergences by adding quarks 
and gluon virtualities $p_j^2$, while using 
dimensional regularization with $d = 4-2\epsilon$ for the UV divergences. As always, the UV divergences are removed
with counterterms, and the IR divergences cancel in the matching. For
renormalization and we will use modified minimal subtraction, with $\mu$ as the $\msbar$ renormalization scale.

The one-loop QCD vertex correction~\cite{Manohar:2003vb} is
\begin{eqnarray}
\cA_\qcd^{(2)} &=& \SetScale{0.5}
  \begin{picture}(90,30) (0,5)
    \SetWidth{0.5}
    \SetColor{Black}
    \GlueArc(90,22.14)(45.86,168.86,11.14){-5}{12.86}
    \ArrowLine(0,31)(90,31)
    \ArrowLine(90,31)(180,31)
    \Photon(86,33)(86,0){5}{5}
  \end{picture}\nn\\
&=&
-\frac{\alpha_s C_F}{4 \pi} 
\left[ 
-\frac{1}{\epsilon} 
+ 2 \log\frac{\pq^2}{\QQ}\log\frac{\pqb^2}{Q^2} 
+ 2\log \frac{\pq^2 \pqb^2}{Q^4} 
+  \log\frac{-Q^2}{\mut^2} 
+  \frac{2\pi^2}{3}
\right]\,.
\end{eqnarray}
We draw the photon explicitly, because this graph depends on choosing $\Gamma=\gamma^\mu$.

The SCET diagrams involve collinear gluons~\cite{SCET1}:
\begin{equation}
\SetScale{0.5}
  \begin{picture}(70,20) (10,10)    
    \SetWidth{0.5}
    \SetColor{Black}
    \ArrowLine(0,6)(60,6)
    \GlueArc(101.36,-0.01)(34.18,161.2,10.13){-5}{8.77}
    \ArrowLine(60,6)(150,6)
    \COval(67,7)(7.07,7.07)(135.0){Black}{White}\Line(63.46,10.54)(70.54,3.46)\Line(70.54,10.54)(63.46,3.46)
    \CArc(101.21,-0.12)(35.17,8.37,156.32)
    \end{picture}
= -\frac{\alpha_s C_F}{4\pi} \left[
-\frac{2}{\epsilon^2} 
-\frac{2}{\epsilon}
+\frac{2}{\epsilon}\log\frac{-p_q^2}{\mut^2}
-\log^2\frac{-p_q^2}{\mut^2}
+2\log\frac{-p_q^2}{\mut^2} -4 + \frac{\pi^2}{6} 
\right] \label{colq}
\end{equation}
\begin{equation}
\SetScale{0.5}
  \begin{picture}(70,20) (10,10) 
\SetWidth{0.5}
    \SetColor{Black}
    \ArrowLine(0,6)(90,6)
    \GlueArc(48.24,2.45)(31.57,171.72,19.52){-5}{8.06}
    \ArrowLine(91,6)(150,6)
    \COval(82,7)(7.07,7.07)(135.0){Black}{White}\Line(78.46,10.54)(85.54,3.46)\Line(85.54,10.54)(78.46,3.46)
    \CArc(48.95,1.84)(33.21,25.23,172.81)
\end{picture}
= -\frac{\alpha_s C_F}{4\pi} \left[
-\frac{2}{\epsilon^2} 
-\frac{2}{\epsilon}
+\frac{2}{\epsilon}\log\frac{-p_\qbar^2}{\mut^2}
-\log^2\frac{-p_\qbar^2}{\mut^2}
+2\log\frac{-p_\qbar^2}{\mut^2} -4 + \frac{\pi^2}{6} 
\right] \label{colqb}
\end{equation}
and soft gluons:
\begin{equation}
\SetScale{0.4}
  \begin{picture}(70,20) (10,10)  
\SetWidth{0.5}
    \SetColor{Black}
    \GlueArc(90,-0.86)(45.86,168.86,11.14){-5}{12.86}
    \ArrowLine(0,8)(90,8)
    \ArrowLine(90,8)(180,8)
    \COval(90,8)(7.81,7.81)(50.194427){Black}{White}\Line(85.76,4.46)(94.24,11.54)\Line(86.46,12.24)(93.54,3.76)
\end{picture}
= -\frac{\alpha_s C_F}{4\pi} \left[
\frac{2}{\epsilon^2} 
-\frac{2}{\epsilon}\log\frac{-p_q^2 p_\qbar^2}{\mut^2 \QQ}
+\log^2\frac{-p_\qbar^2 p_q^2}{\mut^2 \QQ} 
+ \frac{\pi^2}{2} 
\right] \label{soft2}
\end{equation}
We have not assumed the quarks are back to back in computing any of these
diagrams; $n_q \cdot n_\qbar$ appears in the final result only through $\QQ$.
The SCET graphs do not depend on the choice of $\Gamma$. 

The sum of the three SCET diagrams gives
\begin{eqnarray}
\cA_\scet^{(2)} &=&
-\frac{\alpha_s C_F}{4\pi} \left[
-\frac{2}{\epsilon^2} 
-\frac{4}{\epsilon}
+\frac{2}{\epsilon}\log\frac{-\QQ}{\mut^2}
+2 \log \frac{\pq^2}{Q^2} \log \frac{\pqb^2}{Q^2}
+ 2 \log \frac{\pq^2}{Q^2} 
+ 2 \log \frac{\pqb^2}{Q^2} 
\right.\nn\\
&&\left.\hspace{2cm}
- \log^2 \frac{-Q^2}{\mu^2} 
+ 4 \log \frac{-Q^2}{\mu^2}
-8+\frac{5\pi^2}{6}
\right]
\end{eqnarray}
Note how the $\frac{1}{\epsilon}\log p^2_j$ pieces, 
which are UV and IR divergent, drop out of the sum. 
This is a consistency check, as such divergences cannot be removed with
counterterms.

We can see explicitly that the entire dependence on the
IR regulators $p_q^2$ and $p_\qbar^2$
is the same in $\cA_\qcd^{(2)}$ and $\cA_\scet^{(2)}$, as expected. 
Taking the difference between the QCD and the SCET amplitude we find
\begin{equation}
\cA_\qcd^{(2)} - \cA_\scet^{(2)} = 
-\frac{\alpha_s C_F}{4\pi} \left[\frac{2}{\epsilon^2} + \frac{3}{\epsilon} 
- \frac{2}{\epsilon} \log \frac{-Q^2}{\mu^2}
 + \log^2\frac{-\QQ}{\mut^2} -3\log\frac{-\QQ}{\mut^2}+ 8 -\frac{\pi^2}{6}
\right]\,.
\end{equation}
The $1/\epsilon$ UV divergences are canceled by counterterm contributions, and 
matching at $\mut=Q$, we get
\begin{equation}
\label{C2NLO}
\cC_2(Q) = 1 - \frac{\alpha_s C_F}{4\pi} \left[8 -\frac{7\pi^2}{6} +3 \pi i\right]\,.
\end{equation}

To determine the counter terms of the  operator $\cO_2$ we need the wave-function renormalizations as well. 
For the matching, these were not required, since they are exactly
the same in QCD an in SCET~\cite{SCET2}. 
One finds
\begin{eqnarray}
Z_\xi = 1 - \frac{\alpha_s C_F}{4 \pi} \left[ \frac{1}{\epsilon} - \log \frac{-p^2}{\mu^2} + 1 \right]\,.
\label{wavefq}
\end{eqnarray}
Combining this with the vertex diagram gives the result that the UV divergences in the full theory cancel, as is expected for a conserved current. We find that the renormalization constant in QCD is just $1$, while in  SCET
\begin{equation}
 Z_2 = 
1 +\frac{\alpha_s (\mut)C_F}{4\pi} \left[
\frac{2}{\epsilon^2} 
+\frac{2}{\epsilon}\log\frac{-\mut^2}{\QQ}
+\frac{3}{\epsilon}
\right]\,. \label{ZO2}
\end{equation}

\subsubsection{3-jet matching at NLO}

We can also match QCD onto 3-jet operators at NLO. We will not perform
the calculation here, but merely outline what is required and make
some qualitative comments. 
The 3-jet NLO matching involves calculating the difference
between QCD diagrams such as
\begin{equation}
\cA_\qcd^{(3)} =
\SetScale{0.5}
  \begin{picture}(100,40) (0,20) 
 \SetWidth{0.5}
    \SetColor{Black}
   \Gluon(120,75)(180,105){5}{5.98}
    \GlueArc(97.5,37.5)(38.24,168.69,-11.31){-5}{10.29}
    \ArrowLine(15,45)(102,45)
    \ArrowLine(103,45)(180,15)
    \Photon(90,0)(103,44){5}{6}
   \end{picture}
+ \quad \cdots
\end{equation}
and SCET diagrams, such as
\begin{equation}
\cA_\scet^{(3)} \, =\quad
\SetScale{0.6}
  \begin{picture}(110,40) (5,0) 
 \SetWidth{0.5}
    \SetColor{Black}
    \ArrowLine(80,6)(170,6)
    \CArc(106.36,-1)(35,10,156)
    \Line(114,33)(180,50)
    \Gluon(114,33)(181,50){5}{5.23}
    \GlueArc(106.36,-1)(35,156,12){-5}{8.77}
    \ArrowLine(5,6)(65,6)
    \COval(72,6)(7.07,7.07)(135.0){Black}{White}
        \Line(69,10)(75,2)
        \Line(75,10)(69,2)
   \end{picture}
+ \quad \cdots
\end{equation}
There are also contributions from the same diagrams required for $\gamma_3$, below. As for the 2-jet operators at NLO,
the QCD calculation and the SCET calculation are separately divergent, but the divergences all cancel in the difference.

Note, however, that in this case, there is no reason to expect the structure of the operator which results to be the
same as $\cO_3$. So we must add a new operator to the theory 
whose Wilson coefficient starts at $\cO(\alpha_s)$. This operator, when squared (or interfered with $\cO_3$), will produce a differential distribution
for 3-jet events which has a different shape than the tree level 3-parton distribution. In this way, all the shape changes
from loop-corrections in QCD can be reproduced in SCET. We will return to this discussion in 
Section~\ref{subsec:comp:fullQCD}.


\subsection{Running}
While the operators in SCET reproduce the long 
distance physics of QCD, their short distance behavior is completely different. 
In particular, the operator $\cO_2$ is divergent in the UV, while the 
the full QCD current $\qbar \gamma^\mu q$ is not.
The matching performed in the previous section ensured that at some 
particular scale the matrix elements 
in SCET reproduce the matrix elements of QCD exactly at a given order 
in perturbation theory. All the difference between the short distance 
behavior of QCD and SCET above the matching scale $\mu=Q$ is absorbed 
in the precise numerical value of the Wilson coefficients $\cC_n(Q)$. If 
the matching would have been performed at a different scale, the difference 
between QCD and SCET matrix elements would have been different, since the amount of short
distance physics that has to be accounted for is different in that case. Thus,
the Wilson coefficients $\cC_n$ must depend on the value of the matching scale $\mu$. 
Since $\mu$ is just a renormalization scale, it has no observable physical 
effect and our final answer should be independent of this scale. This 
implies that the matrix elements in the effective theory should satisfy 
a renormalization group (RG) equation
\begin{eqnarray}
\mu \frac{d}{d\mu} \left[\cC_2(\mu) \cO_2(\mu) + \cC_3(\mu) \cO_3(\mu) 
+ \ldots + \cC_n(\mu) \cO_n(\mu) \right] = 0\,.
\end{eqnarray}
Since each operator $\cO_n$ contains a different number of labeled collinear fields and 
interactions in the effective theory cannot change this number, 
each contribution has to separately satisfy the RG equation\footnote{
Beyond leading order, there may be mixing among operators. In this case the anomalous dimensions
would be matrices, but we stick to the case without mixing for simplicity.}
\begin{eqnarray}
\mu \frac{d}{d\mu} \left[ \cC_n(\mu) \cO_n(\mu) \right ] = 0\,.
\end{eqnarray}
The $\mu$ dependence enters the operator $\cO_n$ only through its renormalization 
constant $Z_n$. The anomalous dimension of an operator is defined as
\begin{eqnarray}
\gamma_n \equiv \frac{1}{Z_n(\mu)} \mu \frac{d}{d\mu} Z_n(\mu)\,.
\end{eqnarray}
This allows us to write the RG equation in its final form
\begin{eqnarray}
\mu \frac{d}{d\mu} \cC_n(\mu) = \gamma_n(\mu) \cC_n(\mu)\,.
\end{eqnarray}
This differential equation can be written as an integral equation
\begin{eqnarray}
\Pi_n(\mu_2,\mu_1) \equiv \frac{\cC_n(\mu_1)}{\cC_n(\mu_2)} = 
\exp\left[- \int_{\mu_1}^{\mu_2} \frac{d\mu}{\mu}\gamma_n(\mu) \right]\,. \label{evoker}
\end{eqnarray}
We call $\Pi_n$ the RG evolution kernel. It determines the change of a 
Wilson coefficient $\cC_n$ as the scale changes. 
Thus, if we have calculated the Wilson coefficients $\cC_n$ at one scale, we 
can use this kernel to obtain its value at any other scale. 

The anomalous dimensions will have the form
\begin{eqnarray}
\gamma_n = -\left[ \frac{\alpha_s(\mu)}{\pi} \cusp_n^{(1)} + \left(\frac{\alpha_s(\mu)}{\pi}\right)^2 \cusp_n^{(2)} + \ldots \right] \log\frac{\mu^2}{Q^2}-
\left[ \frac{\alpha_s(\mu)}{\pi} \dusp_n^{(1)} + \left(\frac{\alpha_s(\mu)}{\pi}\right)^2 \dusp_n^{(2)} + \ldots \right] \quad .
\label{genform}
\end{eqnarray}
The first term in brackets, is often called the cusp anomalous dimension. It multiplies an explicit,
linear dependence on $\log \mu$. This term arises because of the
double $1/\epsilon^2$ poles in the renormalization constants $Z_n$, which in turn
can be traced back to the fact that full QCD has overlapping soft and collinear 
divergences. Such a term in the anomalous dimension is not problematic if no higher powers of
logarithms appear, because it can be resumed. And in fact, 
it has been shown that at any order in perturbation theory the anomalous dimension contains at most 
a linear dependence on such a logarithm~\cite{Manohar:2003vb}.

As a practical matter, it is helpful to have an explicit form for the evolution kernel~\eqref{evoker}.
At leading order in $\alpha_s$
\begin{equation}
\gamma_n(\mu) = -\frac{\alpha_s(\mu)}{\pi}(\cusp_n \log \frac{\mu^2}{Q^2} + \dusp_n)
\end{equation}
where $\cusp_n$ and $\dusp_n$ do not depend on $\mu$.
The integral over $\mu$ can then be performed explicitly, and we find
\begin{equation}
\Pi_n(Q,\mu) = \exp \left\{
\frac{8\pi}{\beta_0^2 \alpha_s(Q)}
\cusp_n\left(\log\frac{\alpha_s(Q)}{\alpha_s(\mu)}+1-\frac{\alpha_s(Q)}{\alpha_s(\mu)}\right)
-\frac{2}{\beta_0} \dusp_n \log\frac{\alpha_s(Q)}{\alpha_s(\mu)}\right\}
\label{abint}
\end{equation}
From the integrated expression, it is easy to see that the $\dusp_n$ piece is subleading
to $\cusp_n$; the latter has an additional $\log$ enhancement. 
Moreover, when we change the reference scale $\mu_R$ then the coefficient of $\log \alpha_2/\alpha_1$ will shift,
showing that there are additional contributions at NLL of the same order as $\dusp_n$.
Thus, only the cusp anomalous dimension is required for leading log resummation.

We will now determine the anomalous dimensions $\gamma_2$ and $\gamma_3$ at $\cO(\alpha_2)$ and the cusp anomalous
dimension for $\gamma_n$.

\subsubsection{Calculating $\gamma_2$}
The counterterm of the operator $\cO_2$ was already obtained in the previous section
as a byproduct of the calculation of $\cC_2$ at one loop. We found in Eq.~\eqref{ZO2} 
\begin{equation}
\gamma_2(\mu)
= -\frac{\alpha_s(\mu) C_F}{\pi} \left[\log\frac{-\mu^2}{\QQ} + \frac{3}{2}\right]\,. 
\end{equation}
This has the form \eqref{genform} as expected. We can then plug into \eqref{abint} with 
\begin{equation}
\cusp_2=C_F,\quad {\mathrm{and}} \quad
\dusp_2 = C_F \left(\frac{3}{2} + \pi i \right)\,.
\end{equation}
Note that the anomalous dimension is complex. This is due to
a physical effect. Because degrees of freedom have been integrated out, some fields which would go on shell when
we cut a QCD diagram are no longer around.
But for the theories to be the same, the imaginary parts of the cuts must be made up for somewhere else,
and so they show up in the anomalous dimension.

\subsubsection{Calculating $\gamma_3$}
To get the anomalous dimension for $\cO_3$, we need to calculate its renormalization 
constant. We will again regulate the UV by dimensional regularization, and add 
a virtuality to all external particles to regulate the collinear and 
soft IR physics. 
The collinear graphs involving the fermions
are the same as for $\cO_2$, and given in~\eqref{colq} and~\eqref{colqb}. 
The collinear graph involving the gluon is~\cite{Bauer:2001rh}
\begin{equation}
  \SetScale{0.5}
   \begin{picture}(60,30) (20,20)
    \SetWidth{0.5}
    \SetColor{Black}    
    \ArrowLine(93,59)(124,0)
    \ArrowLine(90,63)(159,105)
    \ArrowLine(0,63)(90,63)
    \Gluon(93,59)(125,1){5}{6.92}
    \COval(90,63)(7.07,7.07)(45.0){Black}{White}\Line(86.46,59.46)(93.54,66.54)\Line(86.46,66.54)(93.54,59.46)
    \GlueArc(103.33,40.7)(19.71,111.84,-63.91){-5}{6.69}
    \CArc(100.79,40)(22.32,-53.73,99.78)
   \end{picture}
= 
-\frac{\alpha_s C_A}{4\pi}
\left[
-\frac{2}{\epsilon^2} 
- \frac{1}{\epsilon}
+\frac{2}{\epsilon}\log\frac{-\pg^2}{\mut^2}
-\log^2\frac{-\pg^2}{\mut^2}
+\log\frac{-\pg^2}{\mut^2}
-2 + \frac{\pi^2}{6}
\right]\,. \label{colg}
\end{equation}
The soft graph across the fermions is the same as \eqref{soft2}, up to group theory factors:
\begin{equation}
  \SetScale{0.5}
 \begin{picture}(60,30) (20,20)
   \SetWidth{0.5}
    \SetColor{Black}
    \ArrowLine(92,59)(123,0)
    \Gluon(91,60)(123,2){5}{6.92}
    \ArrowLine(90,63)(159,105)
    \ArrowLine(0,63)(90,63)
    \GlueArc(95.33,55.86)(50.84,171.92,42.19){-5}{11.6}
    \COval(90,63)(7.07,7.07)(45.0){Black}{White}\Line(86.46,59.46)(93.54,66.54)\Line(86.46,66.54)(93.54,59.46)
  \end{picture}
= -\frac{\alpha_s}{4\pi}(\frac{1}{2} C_A -C_F) \left[
-\frac{2}{\epsilon^2}
+ \frac{2}{\epsilon}\log\frac{-\pq^2 \pqb^2}{\mut^2 (\pq + \pqb)^2}
-\log^2\frac{-\pq^2 \pqb^2}{\mut^2 (\pq + \pqb)^2}
-\frac{\pi^2}{2}
\right]
\end{equation}
The soft graph between gluons and quarks are:
\begin{equation}
  \SetScale{0.5}
 \begin{picture}(60,30) (20,20)
   \SetWidth{0.5}
    \SetColor{Black}\ArrowLine(92,59)(123,0)
    \Gluon(91,60)(123,2){5}{6.92}
    \ArrowLine(90,63)(159,105)
    \ArrowLine(0,63)(90,63)
    \COval(90,63)(7.07,7.07)(45.0){Black}{White}\Line(86.46,59.46)(93.54,66.54)\Line(86.46,66.54)(93.54,59.46)
    \GlueArc(104.27,66.5)(37.11,39.28,-79.56){-5}{7.28}
    \end{picture}
= -\frac{\alpha_s C_A}{4\pi} \left[
\frac{1}{\epsilon^2}
-\frac{1}{\epsilon}\log\frac{-\pq^2 \pg^2}{\mut^2 (\pq + \pg)^2}
+\frac{1}{2}\log^2\frac{-\pq^2 \pg^2}{\mut^2 (\pq + \pg)^2}
+\frac{\pi^2}{4}
\right]
\end{equation}
\begin{equation}
  \SetScale{0.5}
 \begin{picture}(60,30)(20,15)
   \SetWidth{0.5}
    \SetColor{Black}
    \ArrowLine(92,59)(123,0)
    \Gluon(91,60)(123,2){5}{6.92}
    \ArrowLine(90,63)(159,105)
    \ArrowLine(0,63)(90,63)
    \COval(90,63)(7.07,7.07)(45.0){Black}{White}\Line(86.46,59.46)(93.54,66.54)\Line(86.46,66.54)(93.54,59.46)
    \GlueArc(82.87,69.23)(48.27,-54.36,-172.59){-5}{9.84}
  \end{picture}
= -\frac{\alpha_s C_A}{4\pi} \left[
\frac{1}{\epsilon^2}
-\frac{1}{\epsilon}\log\frac{-\pqb^2 \pg^2}{\mut^2 (\pqb+\pg)^2}
+\frac{1}{2}\log^2\frac{-\pqb^2 \pg^2}{\mut^2 (\pqb+\pg)^2}
+\frac{\pi^2}{4}
\right]\,.
\end{equation}
Finally, we need the wave function renormalization of the collinear gluon, which is the 
same as in full QCD
\begin{equation}
Z_A=
 -\frac{\alpha_s}{4\pi} \left[\left(\frac{5}{3} C_A - \frac{2 n_f}{3} \right)
\left(-\frac{1}{\eps} + \log\frac{-\pg^2}{\mut^2}\right)\right]\,, \label{wavefg}
\end{equation}
where $n_f$ is the number of flavors.

Adding up the diagrams, the $\frac{1}{\eps} \log p^2$ terms cancel, as they must. 
We find the renormalization constant for $\cO_3$ at one-loop is
\begin{eqnarray}
Z_3  &=& 1 + \frac{\alpha_s}{4 \pi} 
\left[ C_F \left(\frac{2}{\eps^2} 
+ \frac{3}{\eps} 
-\frac{2}{\eps}\log\frac{-(\pq+ \pqb)^2}{\mu^2}\right)\right. \nn \\
&& \hspace{1.5cm}+
\left.
 C_A \left( \frac{1}{\eps^2} 
+ \frac{11}{6\eps}
- \frac{1}{\eps} \log \frac{-(\pq + \pg)^2(\pqb + \pg)^2}{\mu^2 (\pq+ \pqb)^2}\right) 
- \frac{n_f}{3 \eps} \right]\,,
\label{z3exp}
\end{eqnarray}
and the anomalous dimension is
\begin{eqnarray}
\gamma_3 &=& 
 -\frac{\alpha_s}{ \pi} 
\left[C_F \left( \log\frac{-\mu^2}{(\pq+ \pqb)^2} + \frac{3}{2} \right)
+ \frac{C_A}{2} \left( \log \frac{-\mu^2 (\pq+ \pqb)^2}{(\pq + \pg)^2(\pqb + \pg)^2} + \frac{11}{6} \right)
- \frac{n_f}{6}
\right]\,,
\end{eqnarray}
This is of the form \eqref{genform}, with 
\begin{eqnarray}
\cusp_3 &=&
C_F + \frac{1}{2}C_A \\
\dusp_3 &=& 
C_F \left(\frac{3}{2}- \log\frac{-(\pq+ \pqb)^2}{Q^2}  \right)
+ \frac{C_A}{2} \left( \frac{11}{6} - \log \frac{-(\pq + \pg)^2(\pqb + \pg)^2}{Q^2(\pq+ \pqb)^2}  \right)
- \frac{n_f}{6}
\end{eqnarray}
Again, for leading log resummation, only the cusp anomalous dimension $\cusp_3$ is relevant.

\subsubsection{Calculating the leading contribution to $\gamma_n$}
As we have discussed earlier, only the cusp anomalous dimension is required for a LL 
resummation. The cusp anomalous dimension is the coefficient of the $1/\eps^2$ counterterm 
and gets contributions from both soft and collinear diagrams. However, for every $1/\eps^2$ 
divergent term in a particular diagram there is a term proportional to $(\log p^2)/\eps$ which 
cannot be absorbed into a renormalization constant. As we found in the 2-jet matching at NLO, 
this term is canceled once the collinear and soft contributions are added. For the cancellation 
to occur, the total $1/\eps^2$ terms from collinear diagrams have to be 
(-2) times the total $1/\eps^2$ term of the soft diagrams. This implies that the LL contribution
to the anomalous dimension $\gamma_n$ can be obtained from collinear diagrams alone. 

Collinear fields in different directions do not interact with each other at leading order in SCET, thus
the one-loop diagrams from collinear gluons only involve one collinear direction at a time. For
example, the collinear diagrams required for the renormalization of $\cO_2$ were a diagram
involving only the collinear quark (Eqs.~\eqref{colq} and~\eqref{wavefq}), and the antiquark 
(Eqs.~\eqref{colqb} 
and~\eqref{wavefq}). For the operator $\cO_3$ an additional diagrams involving the collinear gluon were
required (Eq.~\eqref{colg} and~\eqref{wavefg}). Note the wavefunction graphs do not have $1/\eps^2$ poles.

Combining these results, $1/\eps^2$ poles in the counterterm of a general operator $\cO_n$ with $n_q$ quark
fields and $n_g$ gluon fields is
\begin{eqnarray}
Z_n = 1 + \frac{\alpha_s}{4 \pi} \left[\frac{n_q C_F +  n_g C_A}{\eps^2}\right]
\end{eqnarray}
which gives the anomalous dimension at LL order
\begin{equation}
\gamma_n = -\frac{\alpha_s}{2 \pi}\left[ n_q C_F + n_g C_A\right] \log \frac{\mu^2}{Q^2}\,.
\end{equation}
Therefore, the cusp anomalous dimension for any operator in SCET is simply
\begin{equation}
\cusp_n =  \frac{n_q}{2}C_F + \frac{n_g}{2} C_A\,.
\end{equation}

\subsection{Threshold matching}
\label{subsec:details:threshold}
Each collinear field from which the operators $\cO_n$ are constructed 
can produce additional particles through interactions described by the Lagrangian 
of SCET. This implies that an operator $\cO_n$ can contribute to final states with more than $n$ partons. For example, we have used in the matching from QCD onto the operator $\cO_3$ that 
the 3-parton matrix element of the operator $\cO_2$ is non-vanishing, and it gave the dominant 
contribution for small values of $p_T$. 
However, in SCET only emissions which keep all external and internal partons near their
mass shell are included, with the amount of virtuality allowed depending on the resolution scale 
of the effective theory. Since virtuality and transverse momentum are related, we use that 
only emissions with $p_T < \mu$ are allowed in SCET. 
This implies that if the scale $\mu$ gets lowered below the $p_T$ of one of the final partons in the 
final state, the original matrix element no longer contributes to the final state, and a threshold
matching onto an operator with an additional collinear field needs to be performed. The threshold
matching condition for the operator $\cO_2$ was already given in Eq.~\eqref{thresholdmatching}, 
and for a general operator $\cO_n^{(j)}$
\begin{equation}
[\cC_n^{(j)} \langle \cO_n^{(j)} \rangle ]_{\mu = p_T+\eps}=
[\cC_{n+1}^{(j)} \langle\cO_{n+1}^{(j)} \rangle ]_{\mu = p_T-\eps}
\end{equation}
We now work out explicitly the matching of  $\cO_2$ onto  $\cO_3^{(2)}$.

The threshold matching of $\cO_2$ onto $\cO_3^{(2)}$ occurs at a scale $p_T$. 
Actually, there are two relevant $p_T$ scales for the branching: the transverse momentum of the gluon with
respect to the quark, $p_T^q$, or to the antiquark, $p_T^\qbar$.  
However, since the SCET results are only valid in the limit $p_T \ll Q$, 
the two $p_T$'s are interchangeable up to power corrections. So we use $p_T = \min(p_T^q, p_T^\qbar)$. We might
also consider $p_T^g$, the transverse momentum the quark (or antiquark) with respect to the gluon. But if $p_T^g$ is 
the smallest transverse momentum, we are well outside of the validity of SCET, and this kinematical configuration is
taken care of by the original matching to QCD. In fact, taking $p_T = \min(p_T^q,p_T^\qbar,p_T^g)$ is equivalent up
to power corrections, and this quantity, to which spherocity reduces for 3-parton kinematics, has the added
property of being infrared safe.

Above the scale $\mu = p_T$, the 3-parton matrix element of $\cO_2$ is  given by the 
sum of the two terms in Eq.~\eqref{O2me}
\begin{eqnarray}
\label{completeO2me}
\langle \cO_2| q \qbar g \rangle
=
g_s\overline{\xi}_{n_q} \left[
\Asl \frac{\nbsl_\qbar}{2} 
 \frac{n_\qbar \tdot( \pq + \pg)}{(\pq+\pg)^2} 
\Gamma
-\Gamma \frac{n_q \tdot( \pqb + \pg)}{(\pqb+\pg)^2} 
 \frac{\nbsl_q}{2} \Asl \right]\xi_{n_\qbar}
\end{eqnarray}
Note that the $1/\bn \tdot (p+q)$ terms in Eq.~\eqref{completeO2me} are due to the non-locality of the 
intermediate quark propagator. However, the propagator scales as $Q/p_T^2$, 
and for $\mu < p_T$ the non-locality of the propagator is less than the resolution of the theory. Thus,
we can think of the matching of $\cO_2$ onto $\cO_3^{(2)}$ at $\mu=p_T$ as serving to keep the theory local 
by adding a label. 

This matrix element can be reproduced by the matrix element of an operator $\cO_3^{(2)}$, defined as
\begin{eqnarray}
\cO_3^{(2)} = g_s 
\overline{\chi}_{n_q}\left[
\cAsl_{n_g} 
\frac{\bnslash_{\bar q}}{2}  \frac{1}{\bn_{\bar q} \tdot {\cal P}^\dagger}\Gamma
-\Gamma
\frac{1}{\bn_q \tdot \cal P} \frac{\bnslash_q}{2} 
\cAsl_{n_g} 
\right]
\chi_{n_\qbar}\,,
\end{eqnarray}
where $\chi_n$ is the same quark jet appearing in $\cO_2$, and 
${\cal A}^\mu_n$ is a gluon jet, that is a collinear gluon field wrapped in Wilson lines, as defined in
\eqref{cadef}.
To satisfy the threshold matching condition we then find that
 the Wilson coefficient $\cC_3^{(2)}$ vanishes for $\mu >p_T$, and for $\mu = p_T$ it is equal to 
the Wilson coefficient of the operator $\cO_2$
\begin{equation}
\cC_3^{(2)}(p_T) = \cC_2(p_T), \quad \cC_3^{2}(\mu > p_T) = 0\,.
\end{equation}
Using previous results, we can write
\begin{equation}
\cC_3^{(2)}(\mu) = \cC_2(Q) \Pi_2(Q,p_T) \Pi_3(p_T,\mu) \Theta(p_T-\mu)
\end{equation}

Threshold matching is just another way of saying that $\cO_2$ emits a gluon at $\mu = p_T$ and turns into $\cO_3^{(2)}$
But it is important to understand the emission process as matching, so that it is improvable.
For example, it would not be hard to do the threshold matching at next-to-leading order. 
This would involve calculating one-loop corrections to both the $\cO_2$ emissions
and to $\cO_2^{(3)}$. The NLO matching
would allow us to go beyond the strongly ordered limit $p_T^1 \gg p_T^2 \gg \cdots \gg p_T^n$, to which the parton
shower is restricted, to correctly describe configurations with  $p_T^1 \gg \cdots \gg p_T^j, p_T^{j+1} \gg \cdots \gg p_T^n$.

\section{Understanding the SCET results}
\label{sec:comp}
In the previous section, we worked out in detail some matching coefficients and evolution kernels in SCET. The important results were summarized
in the beginning of that section.
In this section, we will show how those results can be combined 
to obtain differential jet distributions and show that we agree with traditional perturbative calculations. 
We concentrate on the process $e^+ e^- \to {\rm partons}$, whose kinematics are reviewed in Appendix~\ref{app:kin}.
If we work to order $\alpha_s$ we can have at most three partons in the final state (quark, anti-quark and gluon),
and we can thus obtain the differential decay rate $\rd\sigma/(\rd s \rd t)$.
We begin by showing the Sudakov factors used in parton showers are reproduced in SCET, and discuss the NLL resummation.
We then show that the splitting functions are reproduced in the collinear limit. 
As a corollary, we derive within SCET the classical factorization that parton showers assume. We then
discuss NLO results, and compare to QCD. Finally, we display results from SCET for the thrust distribution and
the 2-jet fraction.

\subsection{Sudakov factors from renormalization group evolution}
\label{subsec:comp:sudakov}
First, we compare the RG evolution kernels obtained in the effective theory
to the Sudakov factors which arise in traditional parton showers. 
We will show that to leading log
accuracy the evolution kernels coincide with traditional Sudakov factors. 

To start, suppose we just match from QCD to 2-jet operators. 
Then, according to \eqref{c2nn},  at $\mu=Q$, $\cC_2^{n,\nbar}=1$ and all other
Wilson coefficients vanish.
For an emission at a scale $p_T$, we need to run $\cC_2$ down to $\mu=p_T$.
Using the 2-jet anomalous dimension
we find
\begin{equation}
\cC_2^{(n,\nbar)}(\mu) = \cC_2(Q) \Pi_2(Q,\mu)\,,
\end{equation}
with the RG Kernel given by
\begin{equation}
\Pi_2(Q,\mu) = 
\exp\left\{
\frac{C_F}{ \pi}\int_{\mu}^{Q} \frac{\rd\mu'}{\mu'}
 \alpha_s(\mu')\left[\log \frac{-\mu'^2}{Q^2} + \frac{3}{2}  \right] \label{pi2Q}
\right\}\,.
\end{equation}

This result can be compared with the expression for the Sudakov factor in traditional parton showers
given in Eq.~\eqref{sud}. The precise form of the Sudakov factor depends on the choice
of evolution variables used, and on the precise value of the argument of $\alpha_s$ that is used. 
Most Sudakov factors use for the scale of  $\alpha_s$ an approximation to transverse momentum given by
\begin{equation}
\tau = z(1-z)t\approx p_T^2\,.
\end{equation}
As an example, we will consider the so-called NLL Sudakov~\cite{Mrenna:2003if,kt,CKKW}, 
which also uses $\tau$ as the evolution variable.
The limits of phase space are $\sqrt{\tau}/Q<z<1-\sqrt{\tau}/Q$.
 Thus the Sudakov factor becomes
\begin{equation}
  \Delta_q^{\mathrm{NLL}} (\tau_2, \tau_1 ) =
 \exp \left\{- \frac{C_F}{2 \pi} 
\int_{\tau_1}^{\tau_2}
  \frac{d \tau'}{\tau'} \alpha_s [ \sqrt{\tau'} ] 
\int_{\frac{\sqrt{\tau'}}{Q}}^{1-\frac{\sqrt{\tau'}}{Q}} \rd z
\frac{1 + z^2}{1 - z} \right\}\,.
\end{equation}
The $z$ integral can be evaluated analytically. Substituting $\mut = \sqrt{\tau'}
\approx p_T$ gives
\begin{equation}
  \Delta_q^{\mathrm{NLL}} (Q, \mut )
=
 \exp \left\{ \frac{C_F}{\pi}
  \int_{\mut}^{Q} \frac{d \mut'}{\mut'} \alpha_s (
  \mut' ) \left[\log \frac{ \mut^2}{Q^2}+\frac{3}{2}  
+ \cO\left(\frac{\mut'}{Q}\right)\right] \right\}\,.
 \label{del2Q}
\end{equation}
Because the integral is dominated for small values of $\mut$, we can
drop the power law $\mut / \mut_0$ terms, as is done in the literature. 
Comparing \eqref{del2Q} to \eqref{pi2Q} we see that the $\cO(\alpha_s)$ evolution kernel reproduces
this Sudakov factor exactly.

The $3/2$ term in \eqref{del2Q} gives rise to subleading logarithms after 
integrating over $\mut$, hence the 
name NLL Sudakov factor. However, there are additional subleading terms which are not 
included consistently. For example, changing the reference scale $\mu_R$ at which the renormalized
 $\alpha_s$ is defined gives rise to 
subleading terms which are also NLL. The optimal value of $\mu_R$ cannot be determined to 
the order we are working. 
Moreover, different Sudakov factors, based on different evolution variables,
give a different constant term~\cite{Mrenna:2003if}. 
In other words, the Sudakov factor only gives the leading logarithms reliably, 
and the $3/2$ term may be dropped.

\begin{figure}[t] 
\begin{center}
 \includegraphics [width=0.8\textwidth,clip]{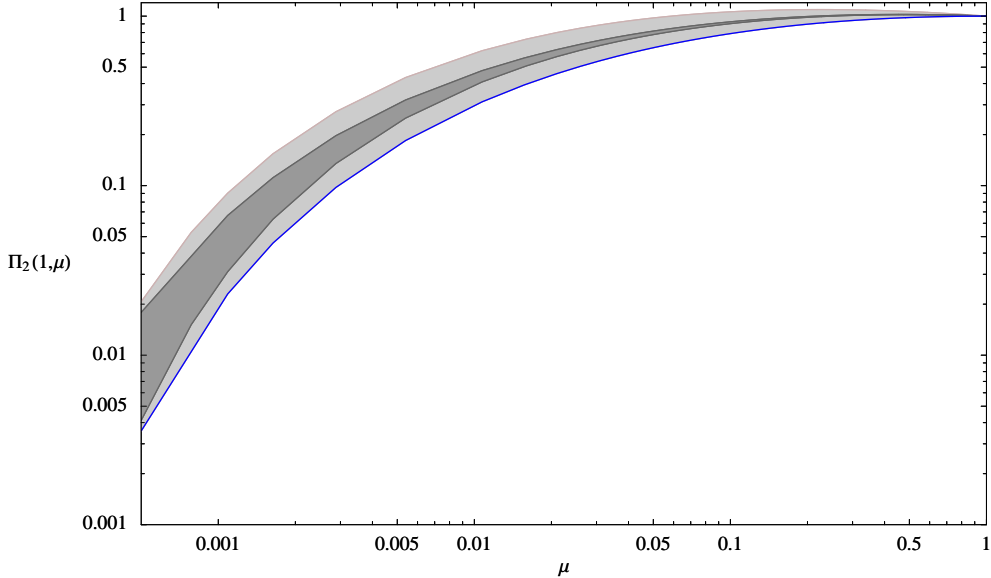}
 \caption{Errors on the RG kernel (Sudakov factor) $\Pi_2$ from next-to-leading log uncertainties.
The light band comes from one NLL effect, varying the $B^1_2 \alpha_s$ term in the anomalous
dimension, from $0<B^1_2<3$. The dark band is from another NLL effect, the 
$\Gamma^2_2 \alpha_s^2 \log $ term in $\gamma_2$, varying $\Gamma^2_2$ between $-2$ and $2$.
We normalize so that $\mu=1$ corresponds to 1 TeV. The LO RG kernel is the lowermost curve
in the figure.
} \label{fig:sudaband}
\end{center}
\end{figure}

To illustrate this point, we show in Figure~\ref{fig:sudaband} the 2-parton
evolution kernel (Sudakov factor)
 $\Pi_2$ with various NLL effects included. 
The light band shows the effect of varying the $3/2$ term 
from $0$ to $3$. The darker band shows the effect of adding a NLL factor proportional to 
$\Gamma_2 \alpha_s^2/(\pi^2) \log \mu'^2/Q^2$ to the integrand 
in Eq.~\eqref{pi2Q}, varying $\Gamma_2$ between $2$ and $-2$
Note that the effect of the
$3/2$ term and the NLL $\Gamma_2$ term are comparable at small $\mu$. 
Since small $\mu$ is precisely
where the Sudakov factors become important, it will be important to include all the NLL resummation
in the Sudakov factors consistently. Later on, we will explore
 the NLL effects on the thrust distribution for 3-parton events (see Figure~\ref{fig:thrust}).

For consistency we will therefore work only at leading-log accuracy, and drop the $3/2$ term. 
At LL all acceptable definitions of the Sudakov factors will give
\begin{eqnarray}
\label{DeltaqLL}
  \Delta_q^{\mathrm{LL}} ( Q, \mut ) &=&
 \exp \left\{  \frac{C_F}{\pi}
  \int_{\mut}^{Q} \frac{d \mut'}{\mut'} \alpha_s (
  \mut' ) \log \frac{\mut'^2}{Q^2}  \right\}\,,
\end{eqnarray}
which is the LL SCET prediction as well.
Analogously, the gluon Sudakov factor (the probability for a gluon not to branch) is
given by~\cite{CKKW}
\begin{eqnarray}
\label{DeltagLL}
  \Delta_g^{\mathrm{LL}} ( Q, \mut ) &=&
 \exp \left\{  \frac{C_A}{\pi}
  \int_{\mut}^Q \frac{d \mu'}{\mu'} \alpha_s (
 \mu' ) \log \frac{\mu'^2}{Q^2}  \right\}\,.
\end{eqnarray}

In SCET, the evolution kernel $\Pi_n$ for an operator with $n$ collinear fields is given by
\begin{eqnarray}
\label{Pin}
\Pi_n(Q,\mu) = \exp\left\{ -\int^Q_{\mu} \frac{d \mu'}{\mu'} \gamma_n(\mu')\right\}\,,
\end{eqnarray}
where $\gamma_n(\mu)$ is the anomalous dimension for the operator, given in Eq.~\eqref{gamman}. 
Comparing Eq.~\eqref{Pin} and~\eqref{gamman} with Eqs.~\eqref{DeltaqLL} and~\eqref{DeltagLL}, 
we find that at LL accuracy the RG 
evolution kernel and the Sudakov factors are related according to
\begin{eqnarray}
\Pi^2_n(Q,\mu) = \Delta_q^{n_q}(Q,\mu)\, \Delta_g^{n_g}(Q,\mu) \,,
\end{eqnarray}
where $n_q$ ($n_g$) are the number of collinear quark (gluon) fields in the operator $\cO_n$. Thus
SCET reproduces all the classical Sudakov no-branching probabilities at leading log through the renormalization group
flow of an effective theory.

\subsection{Splitting functions from collinear emissions}
\label{subsec:comp:splitting}
In this section we will show how 
the couplings in SCET, when put into cross sections, reproduce
the splitting functions of QCD.

For a 3-parton final state, once $\mu$ has run below $p_T$, 2-jet operator can no-longer contribute
and the threshold matching turns $\cO_2$ into   $\cO^{(2)}_3$. If we are not concerned with additional 
emissions, the differential cross section for emission will be
\begin{equation}
\frac{\rd \sigma}{\rd s \rd t} =\frac{\sigma_0}{64\pi^2} \sum_{\mathrm{phys\, pols}} 
\left|\cC_3^{(2)} \langle \cO_3^{(2)}| q \qbar g\rangle\right|^2
\end{equation}
where
\begin{equation}
\sigma_0 = \frac{4 \pi \alpha_e}{3 Q^2} C_A \sum Q_j^2
\end{equation}
We have already seen that $|\cC_3^{(2)}|^2$ encodes the Sudakov factor, so now let us look at the matrix elements.
Using the explicit form of $\cO_3^{(2)}$ given in Eq.~\eqref{o3op}, we 
can perform the sum explicitly. According to the conventions of Section~\ref{sec:details}, we find
\begin{equation}
\left |\langle\cO_3^{(2)}\rangle\right|^2 
= 8 g_s^2 C_F \left[\frac{s}{t} \frac{u^2+Q^2}{(s+t)^2} +  \frac{t}{s} \frac{u^2+Q^2}{(s+t)^2}
+\frac{4 Q^2 u^2}{(t+u)(s+u)(s+t)^2} \right] \label{co32}
\end{equation}

The first term comes from the square of the diagram with the quark emitting, the second from the square
of the diagram with the antiquark emitting, and the third from interference.

Rewriting the amplitude in terms of $t$ and $z$, and taking the limit where the gluon
becomes collinear with the quark, so $p_T^\qbar\to 0$ and $t\to0$, the  amplitude
approaches
\begin{equation}
\frac{\rd \sigma}{\rd s \, \rd t} =  \frac{\sigma_0}{64 \pi^2}
 \left|\langle\cO_3^{(2)}\rangle\right|^2 
= \sigma_0\frac{\alpha_s C_F}{2\pi}\frac{1}{t}\frac{1+z^2}{1-z}+ \cdots
\end{equation}
where the $\cdots$ are higher order in $p_T/Q$.
So we reproduce the QCD splitting function~\eqref{splitfn}, as required.

Note that the interference term (the third term in Eq.~\eqref{co32}) does not have an $s$ or $t$ pole, so
it is finite as $p_T \to 0$ and represents a pure power correction. 
Since the interference is higher order in the SCET expansion, we may simply drop it. 
Recall that dropping interference terms is one of the approximations used in the parton shower, 
and so we see that it is justified by SCET. However, the interference
term should be included following our conventions for evaluating matrix elements -- if we dropped it,
we would not reproduce QCD at the hard scale. Because we match at the matrix element level, it is
important to keep the interference terms in.
Note also that at leading order, we do not need to distinguish $p_T^q$ from $p_T^\qbar$. We only know that
$p_T^q,p_T^\qbar \ll Q$, which is where the SCET amplitude can be trusted.

We can also show that another element of parton showers, namely that successive 
branchings factorize and may be treated classically, can be justified within SCET.
Consider a general operator 
\begin{eqnarray}
\cO = \bar \chi_n \rem
=
  \SetScale{0.6}
 \begin{picture}(100,10) (10,-13)
    \SetWidth{0.5}
    \SetColor{Black}
    \COval(106,-14)(14.87,14.87)(137.72632){Black}{White}
           \Line(98.93,-6.22)(113.07,-21.78)
           \Line(113.78,-6.93)(98.22,-21.07)
    \ArrowLine(90,-14)(31,-14)
  \end{picture}
\end{eqnarray}
where $\rem$ contains any additional collinear fields and the Dirac structure 
of the operator. For example, for the operator $\cO_2$ we would have $\rem = \Gamma \chi_\bn$. 
Now consider the emission of a 
collinear gluon off the collinear fermion $\chi_n$, 
The amplitude for this process is given by
\begin{eqnarray}
A = \bar \chi_n\emi \rem
=
\SetScale{0.6}
\begin{picture}(100,10) (10,0)
    \SetWidth{0.5}
    \SetColor{Black}
    \COval(106,0)(14.87,14.87)(137.72632){Black}{White}
    \Line(98.93,7.78)(113.07,-7.78)\Line(113.78,7.07)(98.22,-7.07)
    \ArrowLine(90,1)(31,2)
    \Gluon(39,29)(77,1){5}{4.29}
    \Line(75,2)(41,28)
  \end{picture}
\end{eqnarray}
where the emission $\emi$ can be obtained from the Feynman rules of SCET. Explicitly,
\begin{eqnarray}
\emi = i g_s T^A\left[\frac{n_{\bar q}^\mu}{n_{\bar q} \cdot p_g} 
+ \frac{1}{\bn_{\bar q} \cdot (p_q+p_g)} \left( \bn_{\bar q}^\mu + \frac{\psl_q^\perp \gamma_\perp^\mu}{n_{\bar q} \cdot p_q}\right)\right]\,.
\end{eqnarray}
The first term comes from the Wilson line emission\footnote{
We include the Wilson line emission here, in contrast
to our previous conventions, to insure gauge invariance. The same results hold if we ignore the Wilson line but
only sum over physical polarizations.}
and the second from the vertex in the SCET Lagrangian.
An important property of $\emi$ is that its square has trivial Dirac structure. In fact,
\begin{eqnarray}
\emidag \emi = 2 g_s^2 C_F \frac{2u(Q^2-s)+t^2}{s\, t\, u} \times {\mathrm{id}}_{4x4}
 \to  8 g_s^2 C_F \frac{1}{z} \frac{1}{t}\frac{1+z^2}{1-z} 
\times {\mathrm{id}}_{4x4}\,.
\end{eqnarray}
The arrow represents the collinear $t\to 0$ limit, 
where we see that QCD splitting function $P(t,z)$ appears (cf. Eq.~\eqref{Pdef}), with
an extra factor of $1/z$.
Now, consider squaring the amplitude $A$ and summing over spins. The spin sum
gives a factor of $\qsl$, which in the collinear limit is $\qsl= z \nsl$, 
which commutes with $\emi$. So,
\begin{equation}
|A|^2 = {\rm Tr} \left\{ \Slash{q} \emi \rem \rem^\dagger \emidag \right\}\nn\\
= z{\rm Tr} \left\{ \emidag \emi \Slash{n} \rem \rem^\dagger \right\}
\end{equation}
Since $\emidag \emi$ 
is proportional to the identity matrix in spinor space, we can pull it  out 
of the trace to obtain the final result
\begin{eqnarray}
|A|^2 = 64 \pi^2 P(t,z) {\rm Tr}\left\{\cO \cO^\dagger \right\}
\end{eqnarray}
This shows that, in the collinear limit, 
the amplitude for a quark to branch into a quark and a gluon is independent of the other
fields in the process, and that the probability for branching is given by a splitting function.

Having considered a single emission of a gluon, we can go one step further and allow for multiple emissions. 
If we call $p_T^{(j)}$ the transverse momentum of the emitted gluon with respect to its 
mother particle, the multiple emissions can be treated as a succession of threshold matchings
if 
\begin{eqnarray}
\label{strongordered}
p_T^{(1)} \gg p_T^{(2)} \gg p_T^{(3)} \gg  \ldots\,.
\end{eqnarray}
This ordering of transverse momenta is called strongly ordered limit. 
In that case, the first emission is encoded in the threshold matching from $\cO_2$ onto 
$\cO_3^{(2)}$, the next emission in the matching from $\cO_3^{(2)}$ onto $\cO_4^{(2)}$, and so  on. 
Since in the above calculation we have assumed a general operator $\bar \chi_n \rem$, the results 
can be applied recursively to the square of the final operator $\cO_n^{(2)}$ that is 
obtained after all the threshold matchings are performed. Thus, the square of the final operator
in the strongly ordered limit can be written as the square of the original operator $\cO_2$, 
multiplied by products of splitting functions. This is precisely the result that a parton shower
algorithm would give for the same process, proven within SCET.

\subsection{Comparison with full QCD perturbative results}
\label{subsec:comp:fullQCD}

In this section, we 
will show how the SCET results produce cross sections which agree with QCD at 
next-to-leading order.
We begin by calculating the most inclusive of quantities, the total cross section for 
$e^+ e^- \to {\rm partons}$, at NLO. This will incorporate the NLO matching
of $\cC_2$ and the 2-jet anomalous dimension. It shows that all the $\cO(\alpha_s)$ 
information from QCD is in
fact contained in the effective theory. Then we consider the differential 
decay rate $\rd \sigma(e^+ e^- \to q \qbar g)$, also at 
NLO~\cite{Ellis:1980wv,Ellis:1980nc,Fabricius:1980fg,Frixione:1995ms,Giele:1991vf,Giele:1993dj,DeRidder:2004tv,Anastasiou:2003gr,Anastasiou:2004qd}.
First, we reproduce the $1/\eps$ and $1/\eps^2$ divergences in this rate
from the counterterms in SCET. Then we show that  at $\cO(\alpha_s)$
all of the large logarithms, which appear in the limit $p_T \ll Q$, are resummed.

We begin with the total cross-section at NLO. 
This total cross section receives contributions 
from both 2- and 3-parton final states, and the 2-parton states have to be calculated 
to one-loop accuracy, while for the 3-parton final states only tree level is 
required~\cite{Kinoshita:1962ur,Lee:1964is}.
In the effective
theory this means that to work consistently at order 
$\alpha_s$ we need the one-loop matching for $\cC_2$ and the tree level matching for $\cC_3$. We
also need the one-loop matrix element for $\cO_2$ and tree level matrix element for $\cO_3$. Both 
the 2- and 3-parton cross sections $\sigma_2$ and $\sigma_3$ are infrared divergent, but 
these infrared divergences cancel in the sum of the two terms.

We begin by calculating the 3-parton cross section $\sigma_3$. In SCET it is obtained 
by squaring the amplitude 
$\cC_2 \langle \cO_2 | q \bar q g\rangle + \cC_3 \langle \cO_3 | q \bar q g\rangle$.
Because of the matching condition~\eqref{match3}, this amplitude is exactly equal to the 
QCD amplitude. Thus, regulating the IR divergence in dimensional regularization,
the 3-jet rate is
\begin{eqnarray}
\sigma_3^{\mathrm{SCET}} &=& 
\int \rd \sep
= 
\sbep \frac{\alpha_s C_F}{2\pi}
 \left( 
\frac{2}{ \eps^2} 
+ \frac{3}{\eps} 
+ \frac{2}{\eps}  \log\frac{\mu^2}{Q^2}  
+\log^2 \frac{\mu^2}{Q^2} + 3 \log\frac{\mu^2}{Q^2}
+\frac{19}{2}
-\frac{7 \pi^2}{6}
\right)\,\,\,,
\end{eqnarray}
where we have written the result in 
terms of the $\mathcal{O} ( \alpha_s^0 )$ 2-parton cross section with dimensionally regulated
phase space
\begin{equation}
  \sbep = \sigma_0\left(\frac{4 \pi}{Q^2}\right)^{\eps} 
\frac{3 ( 1 - \eps )
  \Gamma ( 2 - \eps )}{( 3 - 2 \eps ) \Gamma ( 2 - 2 \eps
  )} \,.
\end{equation}

For the 2-parton cross section we require both the Wilson coefficient $\cC _2$ 
and the matrix element
$\langle \cO_2 | q \bar q \rangle$  at one loop. Since the 3-parton rate above was 
calculated for an arbitrary renormalization scale $\mu$ we 
need the Wilson coefficient at that scale. 
Combining Eqs.~\eqref{C2NLO} with $\Pi_2(Q,\mu)$ we find
\begin{eqnarray}
\cC_2(\mu) = \cC_2(Q)\Pi_2(Q,\mu)
= 1 - \frac{\alpha_s C_F}{4 \pi} \left[ 8 - \frac{\pi^2}{6} 
+ \log^2\frac{-\mu^2}{Q^2}
+3 \log\frac{-\mu^2}{Q^2}
 \right]\,.
\end{eqnarray}
We also need the matrix element of the operator $\cO_2$ at one loop. 
In pure dimensional regularization, 
the one-loop contribution to the bare matrix element vanishes, since in the effective theory
all large scales have been removed from the theory and all infrared scales are set 
to zero. Thus, the only contribution to the matrix element comes from the counterterm, given 
in Eq.~\eqref{ZO2}
\begin{equation}
  Z_2 = 1 + \frac{\alpha_s C_F }{4 \pi} 
 \left[
  \frac{2}{\eps^2} + \frac{3}{\eps} + \frac{2}{\eps}\log \frac{-\mu^2}{Q^2} \right]\,.
\end{equation}
Combining these results, the 2-parton cross section is
\begin{eqnarray}
  \sigma_2^{\mathrm{SCET}} &=& \sbep \left|C_2  \langle \cO_2 |q q  \rangle \right|^2 
= \sbep \left|C_2  \frac{1}{Z_2} \langle \cO_2 |q q  \rangle^{\mathrm{bare}} \right|^2 
\nn\\
&=& \sbep \left[ 1
 - \frac{\alpha_s C_F}{2 \pi}
\left( \frac{2}{\eps^2} 
+ \frac{3}{ \eps} 
-\frac{7 \pi^2}{6} 
+8
+ 2\frac{  \log\frac{\mu^2}{Q^2}  }{\eps}
+\log^2 \frac{\mu^2}{Q^2} 
+ 3 \log\frac{\mu^2}{Q^2} \nn
\right)\right] \,,
\end{eqnarray}
where we have used $\langle \cO_n \rangle = Z_n^{-1} \langle \cO_n\rangle^{\mathrm{bare}}$.

The sum of the 2-parton and 3-parton contributions is finite, and we find for the total 
cross section to order $\alpha_s$
\begin{equation}
 \sigma_{\mathrm{tot}}= \sigma_2^{\mathrm{SCET}} + \sigma_3^{\mathrm{SCET}}
= \sigma_0 ( 1 + \frac{3 \alpha_s}{4 \pi} C_F )\,,
\end{equation}
which is the standard QCD result, reproduced in SCET.

There is an easy way to see why $\sigma_2$ came out the same as in QCD.
 In QCD, the virtual contribution to the 2-parton
cross section is UV finite. Thus, in dim reg, all the $\eps$ dependence comes from IR divergences. 
But the IR
divergences in QCD are the same as in SCET. In SCET, no divergences appear at all when using 
dim reg, because
there are no scales in the problem. Equivalently, in SCET the UV and IR divergences precisely
 cancel. So
the IR divergences are equal to the UV divergences which can be extracted from the counterterm. 
Thus,
the counterterm in SCET has all the information about the full dimensionally-regulated QCD answer,
 up to finite terms. And these finite terms are precisely what is calculated in the matching. 

Just as SCET reproduces the virtual contribution to $\sigma_2$ from QCD,
 it is also capable of reproducing the
NLO contribution to $\rd \sigma_3$. In QCD, this computation involves all 
1-loop contributions to $e^+ e^- \to q q g$.
Again, there are IR divergences, which are canceled when $\sigma_4$, 
the integral of tree-level 4-parton emission,
and the 2-loop contribution to $\sigma_2$ are added. In this paper, we have not included 
all of the relevant computations to reproduce $\sigma_3$ at NLO completely, 
however we have enough information to
reproduce all of the infrared divergences, as well as the dominant large 
logarithms in the finite part.

The next-to-leading order QCD result 
for the dimensionally regularized 3-parton differential cross section 
can be found in~\cite{Ellis:1980wv,Ellis:1980nc}. Consider first the divergent terms. They are
all proportional to the tree-level cross section
\begin{eqnarray}
\rd \sigma_3^{\mathrm{QCD}_1} &=& \rd \sep
 \label{qcdexp} 
\, \frac{ \alpha_s}{2\pi}\, 
\left(\frac{ 4 \pi \mu^2}{Q^2} \right)^{\eps}
\,
 \\
&& \times
 \left\{
-\frac{C_A+2C_F}{\eps^2} 
- \frac{1}{\eps} 
\left[ 3 C_F + \frac{\beta_0}{2} - 2 C_F \log\frac{u}{Q^2} 
- C_A \log\frac{st}{u Q^2} 
\right]
\right\} + \cdots \nn
\end{eqnarray}
The SCET prediction for the 3-parton cross section comes from
\begin{eqnarray}
 \label{scetexp}
\rd \sigma^{\mathrm{SCET}} &=& {\mathrm{PS_3^\eps}} \times
|\cC_2 \Pi_2(Q,p_T) \Pi_3(p_T,\mu)  \langle \cO_3^{(2)} |q \qbar g\rangle
+ \cC_3 \Pi_3(Q,\mu) \langle \cO_3 | q \qbar g \rangle|^2
\\
&=& {\mathrm{PS_3^\eps}} \times
\left|\cC_2 \Pi_2(Q,p_T) \Pi_3(p_T,\mu) \frac{1}{Z_3} \langle \cO_3^{(2)} |q \qbar g\rangle^{\rm bare} 
+ \cC_3 \Pi_3(Q,\mu) \frac{1}{Z_3} \langle \cO_3 | q \qbar g \rangle^{\rm bare}\right|^2\,,\nn
\end{eqnarray}
where $ {\mathrm{PS_3^\eps}}$ refers to the dimensionally regulated 3-parton phases space.
The $1/\eps^2$ and $1/\eps$ poles in QCD come from
soft and/or collinear IR divergences in the matrix elements.
In pure dim reg, the bare matrix elements vanish in SCET, since the UV and IR divergences cancel. Therefore all the poles show up as UV divergences represented by the counterterm. 
Thus, at NLO, the poles are extracted from
Eq.~\eqref{scetexp}  by using $Z_3$ at $\cO(\alpha_s)$
and setting $\Pi_2=\Pi_3=\cC_2=\cC_3=1$.
At order $\cO(\alpha_s)$, the counterterm, from Eq.~\eqref{z3exp}, is
\begin{eqnarray}
Z_3 &=& 1 + \frac{\alpha_s}{4 \pi} 
\left[\frac{2 C_F +C_A}{\eps^2} \right.
\\
&&
\hspace{1.5cm}\left.
+\frac{1}{\eps}
\left(
3C_F+\frac{\beta_0}{2}- 2 C_F \log\frac{-u}{Q^2}-C_A\log\frac{-s t}{u Q^2}
+(2C_F+C_A)\log\frac{-\mu^2}{Q^2}
\right)
\right]\,.\nn
\end{eqnarray}
As expected, $|Z_3^{-1}|^2$ reproduces all of the divergences of the QCD expression.
Note that the matrix elements for $\cO_3$ and $\cO_3^{(2)}$ reproduce QCD, 
by the matching conditions, 
so we get the same $\rd \sep$ factor in both cases. 

The other part of the QCD expression we should be able to reproduce are the dominant large 
logarithms. For the logs to be large, we need $ p_T \ll Q$. In this limit, SCET is
a good approximation to QCD, and $\langle \cO_3| q \qbar g\rangle \approx 0$. Then the large
logs should be resummed in the Wilson coefficients, through the $\Pi_2$
factor. To order $\alpha_s$,
\begin{equation}
\label{Piexp}
\Pi_2(Q,\mu) = 1 - \frac{\alpha_s C_F}{4 \pi} 
\left[
\log^2\frac{\mu^2}{Q^2}+ \left(3+2 i \pi\right) \log\frac{\mu^2}{Q^2} 
\right]\,.
\end{equation}
For $p_T \ll Q$, $\mu=p_T\sim \frac{\sqrt{s t u}}{Q^2}$ (see Appendix~\ref{app:kin}) and the
kinematical structure of the SCET cross section is that of a splitting function. Thus,
\begin{eqnarray}
\frac{\rd \sigma^{\mathrm{SCET}}_3}{\rd s\, \rd t} 
&=& \frac{\sigma_0}{64\pi^2}|\cC_2(Q)\Pi_2(Q,p_T) \langle \cO_2|q \qbar g\rangle|^2+\cdots\nn\\
&=& \sigma_0 P(t,z)
\left[1-\frac{\alpha_s C_F}{2\pi}
\left( 
\log^2\frac{s t u}{Q^6}+ 3 \log\frac{s t u}{Q^6}  
\right)
\right]  + \cdots \label{scetnlo}
\end{eqnarray}

To compare to the QCD result, we need to extract all the
 relevant terms from the NLO 
QCD expression~\cite{Ellis:1980wv}. Since we can only reproduce the logarithmically enhanced 
terms at this order, we will work in the kinematic limit $t \ll s,u$ and expand the full QCD result in 
powers of $t$. 
First, there is a contribution from the pieces with the kinematics of
the  tree-level cross section. This includes the finite piece from Eq.~\eqref{qcdexp},
evaluated at $\mu=s t u$, and an additional finite piece from~\cite{Ellis:1980wv}. All
the terms from these expressions with $\log t$ are: 
\begin{eqnarray}
\frac{\rd \sigma^{\mathrm{QCD}_1}_3}{\rd s\, \rd t} &=&
\sigma_0 
\frac{\alpha_s C_F}{2 \pi} 
\frac{s^2 +t^2 + 2 u Q^2}{s t} \label{qcd1}\\
&& \times \frac{\alpha_s}{2 \pi}
\left[C_F 
\left(\log^2\frac{s t u}{Q^6}-3 \log\frac{s t u}{Q^6}\right)
+\log\frac{s t u}{Q^6}
\left(-\frac{\beta_0}{2} +(2C_F - C_A) \log \frac{u}{Q^2} \right)
\right]\,.\nn
\end{eqnarray}
Next, there is a part of the QCD expression which is not proportional to $\rd \sigma_3$,
but has a splitting function as its collinear limit. Its large logs are
\begin{eqnarray}
\frac{\rd \sigma^{\mathrm{QCD}_2}_3}{\rd s\, \rd t} &=& 
\sigma_0 
\frac{\alpha_s C_F}{2 \pi}
\frac{u^2+(s+u)^2}{s t}
\left[-  (2C_F-C_A) \log \frac{t}{Q^2} \log \frac{u}{Q^2}
\right] \,.\label{qcd2}
\end{eqnarray}
Finally, there is a contribution from the running of $\alpha_s$. The SCET expression
is evaluated with $\alpha_s(\mu)$ while the QCD expression with $\alpha_s(Q)$. Changing
the scale for the QCD expression gives an extra factor:
\begin{equation}
\frac{\rd \sigma^{\mathrm{QCD}_3}_3}{\rd s\, \rd t} 
=\sigma_0 
\frac{\alpha_s(\mu) C_F}{2 \pi} 
\frac{s^2 +t^2 + 2 u Q^2}{s t} \left[\frac{\alpha_s}{2 \pi}
\frac{\beta_0}{2}\log\frac{s t u}{Q^6} \right] \,. \label{qcd3}\\
\end{equation}
Taking the $t \to 0$ limit of these three expressions and adding them we find
\begin{eqnarray}
\frac{\rd \sigma^{\mathrm{QCD}}_3}{\rd s\, \rd t} &=& 
\sigma_0 \frac{\alpha_s C_F}{2 \pi} \frac{1}{t} \frac{1+(1-s)^2}{s} 
\left[ 1 - \frac{\alpha_s C_F}{2 \pi} \left(\log^2\frac{stu}{Q^6} + 3 \log \frac{stu}{Q^6} \right) \right]\,,
\end{eqnarray}
which matches the  SCET expression
in Eq.~\eqref{scetnlo}. Thus SCET resums all the large logarithms.

There are terms in the full QCD expression with logs of $u/Q^2$. In configurations with $u \ll Q$
either the quark and antiquark are back to back, or one of them soft. These configurations
do not come from soft or collinear emission from $\cO_2$, and thus the large logs of $u/Q^2$
are not resummed by $\Pi_2$. However, large logs of $u$ are resummed in $\Pi_3$;
and once SCET incorporates all the 1-loop matching, all of these terms,
as well as all of the finite terms in the QCD expression, should be accounted for.

\subsection{Interpolating between QCD and parton-showers \label{subsec:comp:obs}}
A useful way to explore the SCET prediction is by looking at infrared safe observables. 
We will compare QCD, parton showers, and SCET.
Since we only worked out results for $e^+ e^-$ to 2 or 3 partons, we can only compute observables that depend on
the 3-parton differential distribution. Once SCET is incorporated into a full event generator, more complicated
events can be produced. 
But the results of this section are sufficient to show, at least conceptually, how
SCET interpolates between QCD and the parton shower. So we take our observables to be functions of the 3-parton kinematics:
$f(s,t)$. For example, thrust is
\begin{equation}
T(s,t) = 1-\frac{1}{Q^2}\min\{s,t,1-s-t\}
\end{equation}
Now, let us consider the 3 cases in turn.

First, take tree-level QCD prediction. Thrust is computed as
\begin{equation}
\frac{1}{\sigma_0}\frac{\rd \sigma^{\mathrm{QCD}}}{\rd T} =
 \frac{\alpha_s C_F}{2 \pi} \int \rd s \rd t \frac{s^2 +t^2 + 2 u Q^2}{s t} \delta(T-T(s,t))
\end{equation}
The prediction for QCD at NLO would be the same, but normalized to the
NLO total cross section $\sigma_0(1+\frac{\alpha_s}{\pi})$
instead of just $\sigma_0$.

The parton shower prediction, which resums the leading logs, is given by
\begin{equation}
\frac{1}{\sigma_0}\frac{\rd \sigma^{\mathrm{PS}}}{\rd T} =
 \int \rd s \rd t  \left[\Delta(Q,p_T^q)^2 P(t,z)+\Delta(Q,p_T^\qbar)^2P(s,z')\right] \delta(T-T(s,t))
\end{equation}
where $P(t,z)$ is the splitting function from the quark emission, and $P(s,z')$ is the splitting function from the antiquark emission (here, $z=\frac{Q^2-s}{Q^2+t}$ and $z'=\frac{Q^2-t}{Q^2+s}$).

Finally, the SCET prediction is
\begin{equation}
\frac{1}{\sigma_0}
\frac{\rd \sigma^{\mathrm{SCET}}}{\rd T} = 
\frac{1}{64 \pi^2}
\int \rd s \rd t |\cC_2 \Pi_2(Q,p_T) \langle \cO_3^{(2)}|q \qbar g\rangle
+ \cC_3 \Pi_3(Q,p_T) \langle \cO_3 | q\qbar g \rangle|^2 \delta(T-T(s,t))
\end{equation}
To evaluate this we use the evolution kernels $\Pi_2$ and $\Pi_3$ 
from Eq.~\eqref{pin} with Eq.~(\ref{gamma2},\ref{gamma3}) and the matrix elements. 
The matrix element for $\cO_3^{(2)}$ is given in Eq.~\eqref{co32}: 
\begin{equation}
\left |\langle\cO_3^{(2)}\rangle\right|^2 
= 8 g_s^2 C_F \left[\frac{(s^2+t^2)(u^2+Q^2)}{s t(s+t)^2} 
+\frac{4 Q^2 u^2}{(t+u)(s+u)(s+t)^2} \right] \label{co322}
\end{equation}
The others (evaluated with the conventions of Section~\ref{sec:details}) are:
\begin{equation}
\left|\langle  \cO_3 | q \qbar g \rangle\right|^2 =  16 g_s^2 C_F \frac{u^2(s^2+t^2)}{(t+u)(s+u) (s+t)^2} 
\end{equation}
\begin{equation}
\langle q \qbar g | \cO_3^{(2)}\rangle\langle \cO_3^\dagger| q \qbar g \rangle + {\mathrm{h.c.}}
= -16 g_s^2 C_F \frac{u^2(s^2+t^2)-2 s t u}{(t+u)(s+u) (s+t)^2}  \,.
\end{equation}
Note that the only term that has infrared $s$ or $t$ poles is Eq.~\eqref{co322}.
 The others are finite, and not genuine predictions
of SCET. They depend only on our conventions. 

If it were possible to choose consistent conventions so that 
$\langle \cO_3|q \qbar g \rangle=0$, then
only $\langle \cO_2^{(3)}|q \qbar g\rangle$
would contribute to the distribution, and because of the matching, it would be the same as QCD. In this case,
our distribution would be a Sudakov factor, $\Pi_2^2$ multiplying the QCD cross section, 
which is just the prescription of CKKW~\cite{CKKW}.
So our prediction, at this order, is equivalent to theirs, up to power corrections. Nevertheless, we cannot simply choose 
$\langle \cO_3 |q \qbar g\rangle=0$, because we need a consistent set of conventions which allow us to go to higher orders.

\begin{figure}[t] 
\begin{center}
 \includegraphics [width=0.8\textwidth,clip]{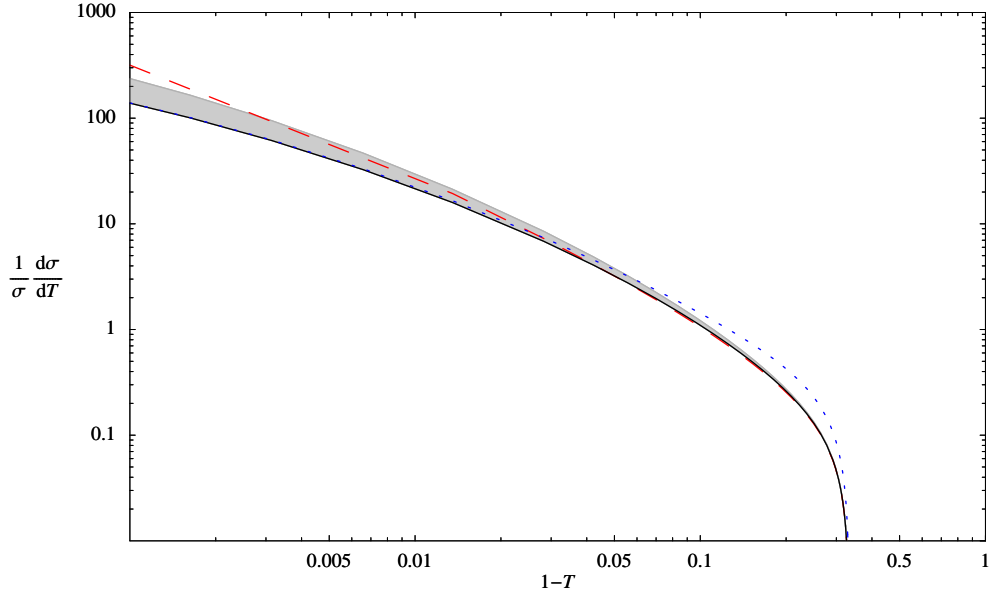}
 \caption{Thrust distribution from 3-parton states, at $E_{\mathrm{CM}}=1$ TeV. 
QCD (dashed red line), parton-shower approximation (dotted blue line), and
SCET (solid black line) are shown. The grey band is a representation of NLL uncertainties,
by varying the $B_2^1$ and $B_3^1$ terms in $\gamma_2$ and $\gamma_3$, between 0 and
their true values.
} \label{fig:thrust}
\end{center}
\end{figure}

The thrust distribution for these three approaches is shown in Figure~\ref{fig:thrust}.
We also include in this figure the NLL uncertainties as a grey band. 
This band corresponds to varying the $\dusp_2$ and $\dusp_3$ terms in the anomalous dimensions
between 0 and their true values. Although the true value is known, this variation
represents the NLL uncertainty from the $\cusp_2^2$ and $\cusp_3^2$ terms, which multiply
$\alpha_s^2 \log \mu$ in $\gamma_2$ and $\gamma_3$, and which is not known. 
Confer Figure~\ref{fig:sudaband} as well.

There are two important features of Figure~\ref{fig:thrust} worth observing. 
First, note how
SCET interpolates between QCD and PS. Small thrust, 
which corresponds
to large $p_T$, is populated by events with hard jets. In this region, we expect QCD
to be a good approximation, as there are no large logarithms, and PS to be a bad approximation,
as we are away from the collinear limit where splitting functions are derived. Note that SCET
approaches QCD in this region. In contrast, large thrust is determined by events with soft or
collinear partons. Here, tree-level QCD is inaccurate, because of large logarithms, while the PS
is closer to reality. SCET matches the parton shower here. So SCET smoothly interpolates between
QCD and PS.

The second feature worth noting is the effect of the NLL resummation. Notice how the grey band is
large for $T\sim 1$, which is where the PS is supposed to be accurate. Thus it seems
the PS is valid neither at small $T$, where hard emissions dominate, nor at large $T$, where
NLL resummation is important. In contrast SCET is valid at small $T$, but also, once
all NLL effects are consistently incorporated, it has the potential to be accurate in all regimes. This shows that
resumming next-to-leading logs may be crucial to get accurate distributions.

Next we compute the 2-jet rate. To do this we need a jet definition~\cite{Sterman:1977wj}.
We will use the the $k_T$ (Durham) algorithm~\cite{kt}.
It defines for any two partons $a$ and $b$
\begin{equation}
y^{a b}\equiv \frac{2}{Q^2} \min\{E_a^2,E_b^2\}(1-cos \theta_{a b}) > \ycut
\end{equation}
We then use
\begin{equation}
y \equiv \min\{ y^{q g}, y^{\qbar g}, y^{q \qbar} \}
\end{equation}
For a 3-parton configuration,
 if $y>\ycut$, there are more than 2-jets, otherwise there are only 2.
Thus we compute the 2-jet rate by integrating the distributions for QCD, SCET, 
and parton showers, over the appropriate range. We avoid the infrared singularities
by integrating over events with $k_T > \ycut$.  For example,
\begin{equation}
\sigma_2^{\mathrm{QCD}}( \ycut) = \sigma_{\mathrm tot}- \frac{\alpha_s C_F}{2\pi}
\int \rd s \rd t \frac{s^2 +t^2 + 2 u Q^2}{s t}
\Theta( y- \ycut)
\label{sig2qcd}
\end{equation}

\begin{figure}[t]
\begin{center}
 \includegraphics [width=0.8\textwidth,clip]{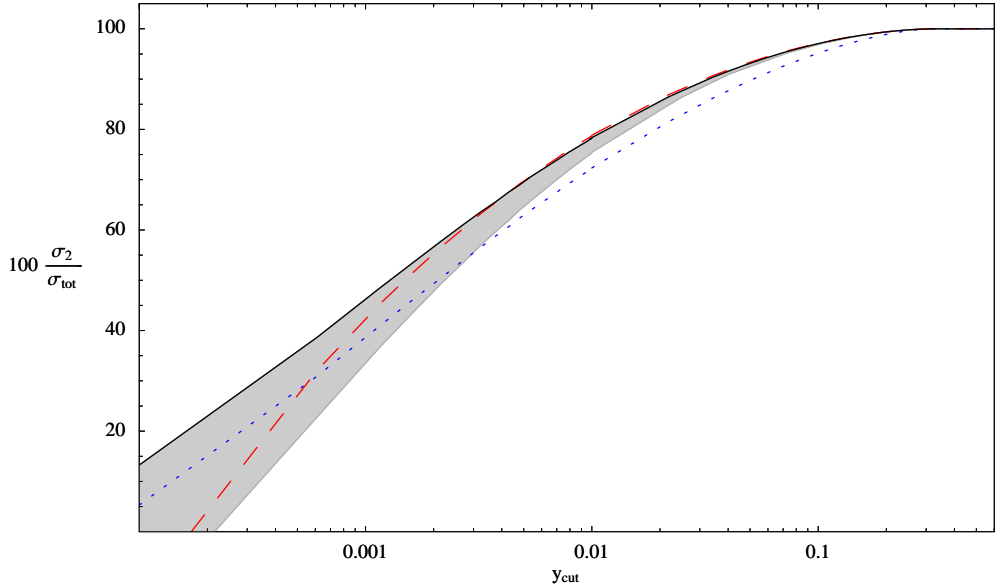}
 \caption{Percentage of events which have 2-jets, as a function of cutoff, using the $k_T$ algorithm.
Shown is QCD (dashed red line), the parton-shower approximation (dotted blue line), and
SCET (solid black line). The grey band is the NLL uncertainty, as in Figure~\ref{fig:thrust}.} 
\label{fig:2jet}
\end{center}
\end{figure}

The results for this observable are shown in Figure~\ref{fig:2jet}.
The effect of the Sudakov suppression can be seen on the left-side, at small $\ycut$. Here, 
the tree-level QCD prediction drops below zero, showing that its estimate of $\sigma_2$ is no
longer trustworthy. Both the PS and the SCET curves are still positive at low energy, implying that
they give a better estimate of the 2-jet cross section. For large $\ycut$, the 2-jet rate should
be accurately given by QCD. The PS gets the rate wrong, because it is integrating over
hard emission where it is not valid. The finite difference between the PS and SCET curves
comes from the integral over large $p_T$, where the PS cannot be trusted. 


\section{Towards SCET event generation \label{sec:algorithm}}
We have shown in the previous section that
SCET reproduces parton showers, Sudakov suppression, and NLO QCD results,
and that it smoothly interpolates between the hard and soft regimes
where QCD and parton showers are valid.
In this section we will illustrate how the SCET formalism is naturally suited to
implementation in an event generator.
It has the capacity to produce particle distributions
with several high $p_T$ jets, while summing the leading logarithms. 
And it is at least as powerful as parton showers for recursively adding 
additional soft or collinear partons.
 
What we are after is the differential cross section $\rd \sigma$ for events
with an arbitrary number of final states. In an event generator, this amounts to
computing the weight, or probability, of an event, 
given the kinematics of the particles in the event.
We will review schematically the analytical expressions
required and give a simple algorithm to obtain events distributed according to 
these distributions.

\subsection{Analytical distributions}
Suppose we want the weight for an event with $r$ particles
from a production process at center of mass energy $Q$. 
To calculate this, the SCET approach requires a sequence of matching and running
steps. First QCD is matched onto SCET at the scale $\mu = Q$ by requiring that 
matrix elements with a given 
number of particles are correctly reproduced by the effective theory. 
In practice, how many particles we include depends on the number of matrix elements that it
is feasible to compute in full QCD (and in SCET). This will in general
be much less than $r$.
Let $m$ be the maximum number of partons which are matched.
So the matching turns on operators $\cO_2$ through $\cO_m$ at $\mu=Q$. 

After the matching, each of the operators is evolved to lower scales 
using the renormalization group equations.
To do this, we need the threshold matching scales, $p_T^{(i)}$, which can be derived
from the event's momenta, for example with the $k_T$ algorithm~\cite{kt}.
So we have $p_T^{(1)}>p_T^{(2)}>\cdots>p_T^{(r)}$. 
At $\mu \sim p_T^{(1)}$ a threshold matching is 
performed where the operator $\cO_2$ is matched onto an operator $\cO_3^{(2)}$.  The set
of operators is then evolved to the scale $\mu \sim p_T^{(2)}$, at which scale 
the operators $\cO_3$ and $\cO_3^{(2)}$ are matched onto 
$\cO_4^{(3)}$ and $\cO_4^{(2)}$. This continues until the scale $\mu$ reaches the
scale $\mu = \ircut$. The differential cross section is then (up to phase space factors)
\begin{eqnarray}
\rd \sigma_r &=& |\sum_{j=2}^m \cC_r^{(j)}(\mu) \cO_r^{(j)}|^2 \label{eventds}\\
&=& \left|[\cC_2(Q) \Pi_2 \Pi_3 \cdots \Pi_r ]\cO_r^{(2)}
+ [\cC_3(Q) \Pi_3 \cdots \Pi_r]\cO_r^{(3)} + \cdots
+ [\cC_m(Q) \Pi_m \cdots \Pi_r] \cO_r^{(m)}\right|^2\,.\nn
\end{eqnarray}
Note that since only operators $\cO_j$ with $j\le m$ were included in the matching 
onto SCET, this expression will only reproduce the full QCD results exactly with up to 
$m$ partons in the final state. All additional partons are described by radiation in SCET
thus rely on an expansion in $p_T$. As we have shown, after squaring the matrix elements, 
the SCET radiation is
equivalent to the splitting functions used in parton showers. So,
\begin{equation}
\rd \sigma_{r>m} = \rd \sigma_m \times {\mathrm{Parton\: Shower}}
\end{equation}
This distribution will be valid if the $p_T$ satisfy
\begin{equation}
p_T^{(1)}, \cdots ,
p_T^{(m)} \gg 
p_T^{(m+1)} \gg \cdots \gg
p_T^{(r)}  
\end{equation}
So if we want the distribution for $r$ hard jets, we should aim for $m \ge r$.

\subsection{A sample algorithm}

Now let us consider how to incorporate these results into an event generator. This means
we need an algorithm for sampling the phase space and unweighting events. 
We do not intend to present a complete or ideal solution, but rather sketch one possibility.
Suppose that the we have QCD and SCET matrix elements for up to $m$ partons,
then an algorithm might look like

\begin{enumerate}
\item Start with a hadronic event at center-of-mass energy $Q$, according to 
as accurate calculations of total cross sections as are available.
\item Pick a configuration of $m$ partons with probability proportional to $\rd \sigma_m$ in
 Eq.~\eqref{eventds}.
\item If the minimum $p_T$ of the configuration, $p_T^{(m-2)}$
 is more than than IR cutoff $\ircut$, then we have an event with $m$ or more partons.
      \begin{itemize}
      \item Start a parton shower at $\mu=p_T^{(m-2)}$, using the SCET RG kernels as Sudakov factors.
      \end{itemize}
\item Otherwise, we have an event with less than  $m$ partons. \label{lessthanm}
    \begin{enumerate}
    \item Pick momenta for $m\!-\!\!1$ partons with probability proportional to $\rd \sigma_{m-1}$ in
Eq.~\eqref{eventds}.
    \item If the minimum $p_T$ of this configuration, $p_T^{(m-3)}$ is more than the IR cutoff $\ircut$,
          then we have an event with $m-1$ partons.
          \begin{itemize}
           \item Do not shower. Go straight to hadronization with this kinematics. 
Or, equivalently, shower starting at $\mu=\ircut$.
          \end{itemize}
    \item Otherwise, we have an event with less than $m-1$ partons. Return to step (a) with $m\to m-1$.
    \end{enumerate}
\end{enumerate}

There are several points worth elaborating on here. 
First, note that the differential distributions in SCET do not diverge as two partons become collinear 
or one becomes soft. 
This is because we are not using the QCD amplitude and correcting with Sudakov factors later on, 
but using the SCET amplitude
which has the Sudakov suppression built in. Since the Sudakov factor vanishes exponentially in 
either of these limits, it overcomes the linear power divergence present in the full QCD 
amplitude.

Second, although we need to use some type of jet definition to sort
the $p_T$ in the kinematics, there will be only a very weak dependence on this 
definition in the final results.
The strongest dependence on the jet definition should cancel from matching to
 the hadronization routine. The detailed
sorting of $p_T$ only has an effect on momenta which are strongly ordered, in
 which case all infrared-safe
jet definitions should agree. There may be a subleading effect, which amounts to 
power corrections, and is beyond the
order we are working.
Also note that because the amplitude is finite even for small $p_T$,
we do not need to employ an intermediate jet cutoff scale in the theory, between $Q$ and $\ircut$.
We emphasize that this algorithm produces fully exclusive events, independent of jet definition. 
For observables with jets, the events can be combined using whatever algorithm is desired, and
 if the same algorithm is
used on the data there should be good agreement.

Third, if we are working to a consistent order in $\alpha_s$,
 the distribution in step \ref{lessthanm} should really be done at NLO for $m-1$ partons, and 
NNLO for $m-2$ partons, {\it etc.}. 
As we have shown, all the NLO information from QCD can
be reproduced in SCET, but one still needs to integrate 
over the regulated phase space for $m$ partons in the singular region
to cancel the NLO divergences for the $m-1$ parton event. 
Quantitatively, the NLO effects are small for the vast majority of observables, so 
this algorithm as presented should be good for most practical purposes.

Finally, the reason we do not just skip step \ref{lessthanm} is as follows. 
The distribution of $m$ partons, from
steps 1 through 3, is as accurate as possible with the information given.
 But because of the Sudakov suppression, the amplitude to produce fewer than
$m$ jets is probably very small. 
In fact, in steps 1 to 3, we make a Monte Carlo estimate of the $m$-or-more parton cross section.
The distribution at small $p_T$ is also sensitive to subleading logarithms,
 and so we probably cannot trust it. Therefore,
step \ref{lessthanm} recalculates the distribution from the hard scale, 
which should be a better estimate of the 
$m-1$ parton event shape.


\section{Conclusion}
We have shown how to construct an event generator based on effective field theory. The correct
effective theory reproducing all collinear and soft divergences of QCD is SCET, and we 
have shown that SCET at leading order
it is equivalent to a conventional parton shower.
The advantage over the parton shower is
that an effective theory is  systematically improvable order-by-order. In particular,
we have shown that by matching SCET onto QCD matrix elements with up to three partons allows 
to obtain cross sections which smoothly merge the parton shower approximation with more accurate
 matrix elements for events with large transverse momentum. 

Since we have a consistent effective field theory framework for the shower, we know precisely
to what order we are working. We can therefore estimate errors, and then incorporate higher
order corrections if necessary when comparing to data. There are five places where
higher order corrections can be incorporated
\begin{enumerate}
\item Matching to QCD for higher multiplicity matrix elements. 
\item Matching to QCD at higher loop order.
\item Higher loop running.
\item Matching within SCET across the emission threshold, at higher order.
\item Power corrections from the SCET expansion.
\end{enumerate}
The first three we have already discussed at length in this paper.
Of these the higher loop running may be the most important to work out and implement.
We showed in Figures~\ref{fig:sudaband} and \ref{fig:thrust} 
an estimate of NLL resummation. It is clearly a large effect for small $p_T$.
Item 4. is not necessarily worth doing. We know that leading order SCET, like the parton shower,
is best when $p_T^{(1)} \gg p_T^{(2)} \gg \cdots \gg p_T^{(n)}$. So if two successive emissions
occur at similar $p_T$'s, the rate will be untrustworthy. However, a large $p_T$ emission will
appear as another jet, and so we really should do the higher order matching to QCD, as in 1.
and 2. to get the rate right. Item 5 is certainly important, but it may be prohibitively difficult to implement.

Our results were collected in the last two sections, and we have indicated how SCET might be used to construct an event generator. We have also investigated the uncertainties in traditional parton showers. For large values of $p_T$, these uncertainties arise from higher order in $p_T/Q$, which are not included due to the approximations required to derive parton showers.
For small values of $p_T$ on the other hand, the subleading logarithms which are not properly summed in the Sudakov factors give rise to large uncertainties as well. So, parton showers have relatively large uncertainties in most regions of phase space. Thus, while
parton showers are very useful to populate phase space of events with many particles, it is very difficult to obtain results with small well defined theoretical uncertainties. Since SCET is able to improve the
precision in all regions of phase space, we showed how SCET might be used to produce events with controllable errors. Our precscription was to
produce an initial $m$-parton distribution at high accuracy, including resummation of large logarithms
within SCET, and then to use the parton shower and hadronization routines, which are constrained by unitary, to fill
out the jets with particles. One can go further, for example by using the properties of SCET, the accuracy of the shower routines could be improved, in principle, to any accuracy desired. Since effective field theories resum logs and are systematically improvable they have the potential to greatly improve our theoretical understanding of particle distributions.

\acknowledgments{
The authors would like to thank  
S.~Ellis, R.~Enberg, S.~Fleming, I.~Hinchliffe, Z.~Ligeti, S.~Mrenna, P.~Skands, and D.~Soper for helpful discussions.
This work was supported by the Director, Office of Science, Office of High Energy and Nuclear Physics, Division of High Energy Physics, of the U.~S.~Department of Energy under Contract DE-AC03-76SF00098 and by a DOE OJI award (CWB).
}

\begin{appendix}

\section{Introduction to SCET}
\label{app:scet}
SCET is an effective theory containing only soft and
collinear degrees of freedom, which can propagate over long distances. Since 
all long distance physics in massless QCD is determined by either soft or collinear
particles, SCET is reproducing the long distance behavior of QCD, while all
short distance physics can be encoded in short distance Wilson coefficients.
SCET is essentially a simplified version of QCD in which all the IR degrees of freedom, that is, the
soft and collinear fields, are the same, but the UV structure is simplified. In particular,
fields which are neither soft nor collinear are integrated out. 

The notation used in SCET emphasizes the nature of the SCET fields. 
Let $n_\mu$ be a lightlike direction $n_\mu = (1,\bfn_i)$, $\bfn^2 = 1$.
Any four-vector can be decomposed  with respect to $n_\mu$ and $\nbar_\mu = (1,-\bfn_i)$ as
\begin{equation}
p^\mu = \frac{1}{2} (\nbar\tdot p) n^\mu + \frac{1}{2} (n \tdot p) \nbar^\mu+ p_\perp^\mu\,.
\end{equation}
So we can define
\begin{equation}
\gamma_\perp^\mu \equiv \gamma^\mu -  \frac{1}{2} \nbsl n^\mu + \frac{1}{2} \nsl \nbar^\mu\,,
\end{equation}
and derive useful identities, such as
\begin{equation}
n \cdot \nbar = 2, \quad
\nsl \nsl=0, \quad 
\{\nbsl,\nsl\}=4, \quad
\{\nsl, \pperpsl\} = 0\,,
\end{equation}
\begin{equation}
p^2 = (n \tdot p)(\nbar\tdot p) + p_\perp^2 \,. \label{pmag}
\end{equation}
We also denote the normalized four-vector in the direction of $p$ by $n_p$. An operator $\cP$ is
often used to project out label momenta. For example,
\begin{equation}
\frac{1}{\nbar\tdot\cP} \xi_n = \frac{1}{\nbar \tdot p} \xi_n\,.
\end{equation}

A field is colliner to $n$ if its momentum satisfies $|p_\perp| < \lambda \, \bn \cdot p$,
where $\lambda$ is a small number giving the expansion parameter in SCET. 
The momentum of a collinear field scales like
$(n\tdot p, \nbar \tdot p, p_\perp) \sim (\lambda^2,1,\lambda)$, where 
we have used \eqref{pmag} with $p^2=0$. A field is
soft\footnote{Sometimes these fields are called
ultrasoft, with soft denoting $p\sim (\lambda,\lambda,\lambda)$. These ``soft'' fields are not
relevant for the current considerations as they cannot interact with either collinear or ultrasoft fields.}
if its momentum scales like $p\sim(\lambda^2,\lambda^2,\lambda^2)$. Note that the sum of two collinear
momenta in the same direction is collinear, so collinear interactions are allowed. But if two fields are collinear to 
different directions, $n_1$ and $n_2$, they scale differently and interactions are forbidden. 

A Dirac fermion $\psi_p$ with momentum $p$ can be decomposed into collinear fermions $\xi_n$ and $\xi_\nbar$ as
\begin{equation}
\psi_p = \frac{\nsl \nbsl}{4} \psi_p + \frac{\nbsl \nsl}{4} \psi_p \equiv \xi_n + \xi_\nbar\,.
\end{equation}
The difference betwen $\psi_p$ and $\xi_n$ is that on-shell, $\psi_p$ satisfies 
$\psl \psi_p=0$ while $\xi_n$ satisfies $\nsl \xi_n=0$. When $n$ is aligned with $p$,
$\psi_p=\xi_{n_p}$, but in general $\psl \xi_n \ne 0$.
The SCET Lagrangian is derived from the QCD Lagrangian by integrating out 
the small components $\xi_\nbar$ \cite{SCET1}.  Then, using the scaling properties
of the quarks and soft and collienar gluons, the Feynman rules are worked out
as an expansion in $\lambda$. They can be found in \cite{SCET1,SCET2,SCET3,SCET4}.

SCET has the curious property that no information is lost when $\xi_\nbar$ is integrated out,
even if only the
first order terms in the $\lambda$ expansion are kept. There are a few ways to see this. First,
note that $(\nbsl \nsl)/4$ and $(\nsl \nbsl)/4$ are projectors, in that they are orthogonal and
complete. However, we can regain $\psi_p$ from just $\xi_n$ or $\xi_\nbar$
using the identity
\begin{equation}
\xi_{n_1} = (1+\frac{\pperpsl \nbsl_1}{2 \nbar_1 \tdot p} ) \xi_{n_2} \,,\label{diractojet}
\end{equation}
and that  $\psi_p=\xi_{n_p}$.
One should keep in mind that SCET is a 
boosted version of QCD~\cite{SCET1}, and any interactions
between collinear fiels in the same directions are thus given by full QCD. 
Whenever there are fields collinear to different directions, however,
their interactions will be different from in QCD. It is in this situation that SCET is useful.

Since collinear fields in different directions do not interact with one another, there should 
be a separate collinear gauge invariance for each direction $n$. To ensure this gauge invariance, 
however, requires that each collinear fermion is multiplied by a collinear Wilson line, making the 
resulting field gauge invariant by itself
\begin{equation}
\chi_n = W_n \xi_n,\quad W_n = \exp\left\{g_s \frac{\nbar \tdot A}{\nbar\tdot p_A} \right\}\,.
\end{equation}
Each collinear fermion is therefore wrapped in collinear Wilson lines, and it is these collinear "jets" which are the basic building blocks of operators in SCET. For example, an operator with two jets would be
\begin{equation}
\cO_2^{(n_1, n_2)} = {\bar \chi}_{n_1} \Gamma \chi_{n_2}\,,
\end{equation}
where $\Gamma$ is some tensor structure. Because $n_1$ and $n_2$ are different labels, no collinear fields can couple
to both jets. Collinear gluons can couple one jet to itself, and soft gluons can be exchanged between jets. 

Diagrams in SCET are usually computed in $\msbar$ dimensional regularization. It is also helpful
to regulate IR divergences by giving the quarks and gluons small virtualities.
These virtualities must drop out of physical
calculations, and it is a helpful check on the theory to show that they do. 
Sample calculations can be found in~\cite{SCET3}. Physical quark masses can easily be accounted for
in SCET~\cite{SCETmasses}, but for simplicity we take all fields to be massless.

\section{3-parton Kinematics \label{app:kin}}
For the
three parton final state, label the momenta of the quark, antiquark and gluon $\pq^\mu$, $\pqb^\mu$ and $\pg^\mu$, 
repsectively. There are two independent invariants we can construct. However, it is convient to
go between a number of kinematic variables. Let $Q$ be the center-of-mass (COM) energy of $e^+ e^-$.
Then, we have
\begin{eqnarray}
\label{studefs}
s &\equiv& (\pqb + \pg)^2 \equiv Q^2(1 - x_q )\nn\\
t &\equiv& (\pg + \pq)^2 \equiv Q^2(1 - x_\qbar )\nn\\
u &\equiv& (\pq + \pqb)^2 \equiv Q^2(1 - x_g) \,.
\end{eqnarray}
The invariants satisfy $s+t+u=Q^2$ and $x_q +x_\qbar + x_g = 2$. The $x_q$ are half the energy of the corresponding
particle, in the COM frame. 
We also define the four light-like vectors pointing in the directions of the three particles as $n_q$,
$n_\qbar$ and $n_g$. So
\begin{eqnarray}
\pq^\mu = \frac{Q}{2} x_q  n_q^\mu \,, \qquad 
\pqb^\mu = \frac{Q}{2}x_\qbar n_\qbar^\mu \,, \qquad 
\pg = \frac{Q}{2} x_g n_g^\mu\,.
\end{eqnarray}
For each particle, we will also need the light-like vectors pointing in the 
opposite direction 
\begin{eqnarray}
n_i = (1, \bfn_i) \quad \Rightarrow \quad \bar n_i = (1,-\bfn_i)
\end{eqnarray}
The scalar products between different $n_i$ can be obained
from \eqref{studefs}, and we also have 
\begin{eqnarray}
\bar n_i \tdot \bar n_j = n_i \tdot n_j \,, \qquad 
\bar n_i \tdot n_j = 2 - n_i \tdot n_j\,.
\end{eqnarray}
We can also derive
\begin{equation}
\overline{n_\qbar} \tdot( \pq + \pg) = \frac{t}{Q}, \qquad n_\qbar \tdot(\pq +\pg) = Q
\end{equation}
and its permutations.

In the COM frame, 
the transverse momentum of the $\pg$ or $\pqb$ with respect to $\pq$,
 $\pq$ or $\pqb$ with respect to $\pg$, and $\pqb$ or $\pg$ with respect to $\pqb$
are given, respectively, by
\begin{eqnarray}
p_T^q = \frac{\sqrt{s t u}}{x_q Q^2} \,, \qquad 
p_T^g = \frac{\sqrt{s t u}}{ x_g Q^2 }\,. \qquad
p_T^\qbar = \frac{\sqrt{s t u}}{ x_\qbar Q^2}\,.
\end{eqnarray}
We will also define $p_T$ as
\begin{eqnarray}
\label{pTdef}
p_T = {\rm min}(p_T^q, p_T^\qbar, p_T^g)\,.
\end{eqnarray}
Note that for $p_T \ll Q$,
\begin{equation}
p_T^2 \approx \frac {s t u}{Q^4}\,.
\end{equation} 
Finally, it is helpful when discussing splitting functions, to use variables defined with respect to only two
partons. If we are considering the quark and gluon, then we can use the invariant mass of the pair, and the
energy fraction of the gluon
\begin{equation}
t = Q^2(1-x_\qbar), \qquad z = \frac{x_q}{x_g+x_q}\,.
\end{equation}
Some useful relations between these variables and the others are
\begin{equation}
x_\qbar = 1-\frac{t}{Q^2}, \qquad x_q = z(1+\frac{t}{Q^2}), \qquad x_g = (1+\frac{t}{Q^2})(1-z)\,.
\end{equation}
\end{appendix}

\end{document}